\documentclass[useAMS,usenatbib]{mn2e}
\usepackage{graphicx}

\title[3D Spectroscopy of Local Luminous Compact Blue Galaxies]{3D Spectroscopy of Local Luminous Compact Blue Galaxies: Kinematic Maps of a Sample of 22 Objects}
\author[J. P\'erez-Gallego, et al.]{J. P\'erez-Gallego$^{1,2}$\thanks{E-mail:jgallego@astro.ufl.edu}, R. Guzm\'an$^{1}$, A. Castillo-Morales$^{2}$, J. Gallego$^{2}$, F. J. Castander$^{3}$,
\newauthor
  C. A. Garland$^{4}$, N. Gruel$^{7,1}$, D. J. Pisano$^{5,6}$ and J. Zamorano$^{2}$\\
$^{1}$University of Florida, 211 Bryant Space Science Center, Gainesville, FL 32611, USA. \\
$^{2}$Departamento de Astrof\'isica y CC. de la Atm\'osfera, Universidad Complutense de Madrid, Madrid, Spain.\\
$^{3}$Institut de Ci\`encies de l'Espai (IEEC/CSIC), Campus UAB, 08193 Bellaterra, Barcelona, Spain\\
$^{4}$Department of Natural Sciences, Black Science center, Castleton College, Castleton, VT 05735, USA.\\
$^{5}$West Virginia University, Department of Physics, P.O. Box 6315, Morgantown, WV 26505, USA.\\
$^{6}$Adjunct Assistant Astronomer at the National Radio Astronomy Observatory.\\
$^{7}$Centro de Estudios de F\'isica del Cosmos de Arag\'on, 44001 Teruel, Spain.\\
}
\begin{document}

\date{Accepted 2011 August 11. Received 2011 August 5; in original form 2010 June 9}

\pagerange{\pageref{firstpage}--\pageref{lastpage}} \pubyear{2010}

\maketitle

\label{firstpage}

\begin{abstract}
We use three dimensional optical spectroscopy observations of a sample of 22 local Luminous Compact Blue Galaxies (LCBGs) to create kinematic maps. By means of these,
we classify the kinematics of these galaxies into three different classes: rotating disk (RD), perturbed rotation (PR), and complex kinematics (CK). We find 48\% are RDs, 28\% are PRs, and 24\% are CKs. RDs show rotational velocities that range between $\sim50$ and $\sim200~km~s^{-1}$, and dynamical masses that range between $\sim1\times10^{9}$ and $\sim3\times10^{10}~M_{\odot}$.
We also address the following two fundamental questions through the study of the kinematic maps: 
\emph{(i) What processes are triggering the current starbust in LCBGs?} 
We search our maps of the galaxy velocity fields for signatures of recent interactions and close companions that may be responsible for the enhanced star formation in our sample.
We find 5\% of objects show evidence of a recent major merger, 10\% of a minor merger, and 45\% of a companion. This argues in favor of ongoing interactions with close companions as a mechanism for the enhanced star formation activity in these galaxies. 
\emph{(ii) What processes may eventually quench the current starbust in LCBGs?} Velocity and velocity width maps, together with emission line ratio maps, can reveal signatures of Active Galactic Nuclei (AGN) activity or supernova (SN) driven galactic winds that could halt the current burst. We find only 5\% of objects with clear evidence of AGN activity, and 27\% with kinematics consistent with SN-driven galactic winds. Therefore, a different mechanism may be responsible for quenching the star formation in LCBGs.
Finally, from our analysis, we find that the velocity widths of RDs, rather than accounting exclusively for the rotational nature of these objects, may account as well for other kinematic components, and may not be good tracers of their dynamical masses. 
\end{abstract}


\section{Introduction}\label{introduction}

Luminous Compact Blue Galaxies (LCBGs) are small, but vigorously star forming galaxies.  The star formation rates (SFRs) per unit mass are high and
the star formation activity involves up to $\sim10\%$ of the total galaxy mass \citep{koo94,guzman96,guzman97,guzman98,phillips97,gil00,hammer01,noeske06,puech06,rawat07}.  LCBGs are a major contributor to the cosmic SFR increase in the last $\sim9$ Gyr \citep{lilly98,mallen99,melbourne07} and are related to distant Lyman-break galaxies \citep[LBGs;][]{steidel96,lowenthal97}, the brightest blue compact dwarfs \citep[BCDs;][]{cairos01}, local HII and starburst nuclei galaxies \citep[SBNs;][]{werk04}, and intermediate-redshift compact narrow emission-line galaxies \citep[CNELGs;][]{koo94}. Their presence at different redshifts denotes their cosmological relevance and implies that local starburst galaxies, when properly selected, are unique laboratories for studying the complex ecosystem of the star formation process over time.

Two key questions regarding the nature of LCBGs remain unanswered: (i) What is the role of mergers or other interactions in triggering the burst of star formation?; and (ii) What is the role of Active Galactic Nuclei (AGN) and supernova (SN) driven galactic winds in quenching the star formation and limiting the stellar mass?

Previous results show that while approximately 50\% of local LCBGs are not isolated and have close optical 
companions \citep[some gas-rich companions are only revealed by means of neutral hydrogen (HI) observations; e.g., Pisano et al. in preparation,][]{garland04}, only $\sim30\%$ of field galaxies are not isolated \citep{james08,pisano02}. A correlation between the kinematic signatures of recent interactions in LCBGs and their environment would imply that these interactions are responsible for triggering the enhanced star formation in LCBGs.

Kinematic signatures of recent interactions should be easy to identify by means of velocity and velocity width maps. In particular, major and minor mergers cause deviations from smooth rotation, which can be identified with high resolution data. As an example, \citet{homeier99} studied whether a minor or a major merger in NGC 7673 was responsible for triggering the enhanced star formation in this particular galaxy. They, as well as  \citet{perez09}, concluded that while its asymmetric outer features point to a major merger, the inner disk structure constrains the strength of the event to the scale of a minor one. We extend the study of \citet{perez09} to include 22 objects in this paper.

Another interesting feature of LCBGs is that their typical stellar masses \citep[i.e., from $\sim5\times10^{9}$ to $\sim10^{10}~M_{\odot}$;][]{gil00,guzman03}, place them at the location of the upper limit of galaxies belonging to the ``blue cloud" \citep[$M_*\sim3\times10^{10}~M_{\odot}$;][]{kauffmann03,blanton03}. Their stellar masses turn LCBGs into unique laboratories to investigate why the star formation and the stellar mass growth stop in blue galaxies. Quenching mechanisms include massive halos, AGNs, and SN-driven galactic winds. Halo mass measurements are needed to study the effects of massive halos, while both flux and kinematic signatures are needed to study the presence of AGN and SN driven galactic winds in these galaxies.

Halo mass estimates can be inferred from HI dynamical masses. Instead of ionized gas, neutral atomic gas is used because it traces the gravitational potential to larger radii. Typical HI dynamical mass estimates of LCBGs are between $10^{10}~M_{\odot}$ and $10^{12}~M_{\odot}$ as shown by \citet{garland04} and Pisano et al. (in preparation).
In addition, by comparing resolved HI dynamical masses with dynamical masses derived from our optical analysis, we can infer the actual dynamical masses of distant LCBGs, where we can not measure neutral atomic gas fluxes.

Other quenching mechanisms include AGNs and SN-driven galactic winds \citep{mihos94,mihos96,sanders96,granato04,murray05,springel05,hopkins06}. 
While the presence of quasars can be ruled out after investigating the Sloan Digital Sky Survey (SDSS) spectra of these galaxies, we cannot rule out the presence of faint AGNs in LCBGs \citep[e.g.,][]{cairos09}. In particular, optical Integral Field Spectroscopy (IFS) has the potential of revealing the presence of low-level AGN by means of unusually high line ratios [OIII]/H$\beta$ and [NII]/H$\alpha$ \citep{brinchmann04}, and spectrally resolved kinematic components with broad Full Widths at Half Maximum (FWHMs). Furthermore, the extent and strength of such kinematic subcomponents allows us to quantify the outflow/inflow of gas from/onto galaxies by means of the study of their line profiles\citep{marlowe95}. It is important to note that IFS can derive these properties as a function of position within the galaxy.

In addition, we can use local LCBGs as a template for studies of distant blue starbursts at z$\geq$1.
Recent surveys aim to characterize the spatially resolved kinematics of distant (i.e., from $z\sim0.4$ to $z\sim2$) starbursts by means of IFS using instruments such as GIRAFFE and SINFONI at the Very Large Telescope of the European Southern Observatory \citep[e.g.,][]{epinat09,puech06,forster06}. Nearby LCBGs can be used as proxies for learning about the formation, nature, and evolution of distant starbursts, and to simulate observations of those in order to help us properly interpret the results of current and future distant surveys (Gruel et al. in preparation).

In \citet{perez09} and \citet{castillo10} we showed the analysis of prototypical LCBG NGC 7673. NGC 7673 is the first of a sample of 22 LCBGs within approximately 100 Mpc selected from the SDSS, Universidad Complutense de Madrid \citep[UCM;][]{zamorano94,vitores96,gallego97,gonzalez03}, and Markarian \citep{markarian89} catalogs.

In this paper we focus on the optical kinematic properties of a sample of 22 LCBGs, and show their three dimensional optical spectroscopy observations. We use these to shed light on the questions highlighted above. Flux related properties, such as SFRs, metallicities, and extinctions, are discussed in Castillo-Morales et al. (in preparation).

Throughout this paper we assume a flat universe with $\Omega=0.7$, $\Lambda=0.3$, and $h=0.7$. The paper is structured as follows: in Section 2 we describe our sample selection, observations, data reduction, and measurements; the analysis and discussion are carried on in Section 3; and finally, in Section 4, we summarize our work.

\section{Sample Selection, Observations, Data Reduction, and Basic Measurements}\label{data}

\subsection{Sample Selection}\label{selection}

LCBGs are starburst galaxies which form a continuous distribution in luminosity, surface brightness, and color.  Our observational selection criteria \citep{pisano01,garland04,werk04} select objects with (i) absolute blue magnitude ($M_B$) brighter than $-18.5$~mag; (ii) effective surface brightness ($SBe$) brighter than 21 $B$-mag~arcsec$^{-2}$; and (iii) rest-frame $B-V$ color bluer than 0.6~mag. These criteria ensure similarity between nearby LCBGs and distant blue starbursts seen in wide-field Hubble Space Telescope imaging surveys \citep{scoville07}.

We selected 20 local ($D<200$~Mpc) LCBGs from the Data Release 4 \citep[DR4;][]{adelman06} of the SDSS and UCM catalogs. We completed our sample with NGC 7714, and SDSS 1703b (companion to SDSS 1703a, with properties within 1-$\sigma$ of our selection criteria) for a total of 22 objects (Table~\ref{table:sample}). Note that these sources cover the whole range of $M_B$, $SB_e(B)$, and $B-V$ of the complete sample of local LCBGs in the DR4 of the SDSS; their properties are listed in Table~\ref{table:properties}. For this purpose we refer to our sample as representative of the LCBG population. LCBGs with effective radii larger than 4~arcsec and closer than 100~Mpc were given priority to take full advantage of the field of view (FOV) and spatial resolution of the instrument used. 

\begin{table*}
\centering
\caption{Sample$^{\mathsf{a}}$}
\vspace{0.5cm}
\begin{tabular}{lccccl}
\hline
Name 		& Alternate Names	 & RA (2000)   & DEC (2000) & $z$$^{\mathsf{b}}$ \\
\hline		
NGC 7673	& MRK 325, IV ZW 149	 & 23h27m41.0s & +23d35m20s & 0.011368 \\
NGC 7714 	& MRK 538, ARP 284	 & 23h36m14.1s & +02d09m19s & 0.009333 \\
NGC 6052 	& MRK 297, ARP 209 	 & 16h05m12.9s & +20d32m32s & 0.015808 \\
NGC 469   	& \ldots		 & 01h19m32.9s & +14d52m19s & 0.013673 \\
UCM 0000+2140 	& MRK 334, IV ZW 1	 & 00h03m09.6s & +21d57m37s & 0.021960 \\
UCM 0156+2410	& \ldots		 & 01h59m15.7s & +24d25m00s & 0.013030 \\
UCM 1428+2727	& MRK 685, HARO 42	 & 14h31m08.9s & +27d14m12s & 0.014890 \\
UCM 1431+2854	& \ldots		 & 14h33m20.7s & +28d41m36s & 0.031000 \\
UCM 1648+2855	& MRK 1108		 & 16h50m47.9s & +28d50m45s & 0.032969 \\
UCM 2250+2427	& MRK 309, IV Zw 121	 & 22h52m34.7s & +24d43m50s & 0.042149 \\
UCM 2258+1920	& \ldots		 & 23h01m07.1s & +19d36m33s & 0.021592 \\
UCM 2317+2356	& NGC 7620, MRK 321	 & 23h20m05.7s & +24d13m16s & 0.031962 \\
UCM 2327+2515n	& IV ZW 153		 & 23h30m09.9s & +25d31m58s & 0.020600 \\
UCM 2327+2515s	& \ldots		 & 23h30m09.9s & +25d31m58s & 0.019130 \\
SDSS 1134+1539	& \ldots		 & 11h34m21.2s & +15d39m37s & 0.017335 \\
SDSS 1507+5511	& \ldots		 & 15h07m48.4s & +55d11m08s & 0.011231 \\
SDSS 1605+4120	& MRK 1104		 & 16h05m45.9s & +41d20m41s & 0.006640 \\
SDSS 1652+6306	& MRK 1109		 & 16h52m03.6s & +63d06m57s & 0.010376 \\
SDSS 1703+6127a	& \ldots		 & 17h03m14.9s & +61d27m04s & 0.019690 \\
SDSS 1703+6127b	& \ldots		 & 17h03m12.2s & +61d27m21s & 0.019760 \\
SDSS 1710+2138	& \ldots		 & 17h10m11.1s & +21d38m58s & 0.014757 \\
SDSS 2327-0923	& NGC 7675		 & 23h27m14.8s & -09d23m13s & 0.016044 \\
\hline
\multicolumn{5}{l}{$^{\mathsf{a}}$ Throughout the paper these galaxies are referred to by a truncation of the name listed here.} \\
\multicolumn{5}{l}{$^{\mathsf{b}}$ The typical uncertainty for $z$ is $\pm$0.000025 for galaxies in the SDSS and UCM catalogs.}\\ 
\end{tabular}
\label{table:sample}
\end{table*}

\begin{table*}
\centering
\caption{Sample Properties$^{\mathsf{a}}$}
\vspace{0.5cm}
\begin{tabular}{lcccccccc}
\hline
Name & $z$$^{\mathsf{b}}$ & $d_{a}$$^{\mathsf{c}}$ & $M$($B$)$^{\mathsf{c}}$ & $m$($b$)$^{\mathsf{c}}$ & $SB_{e}$$^{\mathsf{c}}$ & $B-V$$^{\mathsf{c}}$ & $R_e$$^{\mathsf{c}}$ & $R_e$$^{\mathsf{c}}$\\
{} & {} & (Mpc) & (mag) & (mag) & (B-mag arcsec$^{-2}$) & (mag) & (arcsec) & (kpc)\\
\hline	
NGC 7673$^{\mathsf{e}}$      		& 0.011368 	& 48.0     	& -20.36     & 13.28     & 19.95  	& 0.41  & 8.9     & 2.1 \\
NGC 7714			 	& 0.009333 	& 39.7     	& -20.10     & 13.00     & 20.00  	& 0.40  & 14.3    & 2.7 \\
NGC 6052$^{\mathsf{d}}$      		& 0.015808 	& 67.6     	& -20.69     & 13.37     & 19.71  	& 0.41  & 7.6     & 2.4 \\
NGC 469$^{\mathsf{d}}$   		& 0.013673 	& 58.4     	& -19.13     & 14.74     & 20.05  	& 0.42  & 4.7     & 1.3 \\
UCM 0000+2140     			& 0.021960 	& 91.6    	& -20.50     & 14.61     & 20.87  	& 0.30  & 7.5     & 3.7\\
UCM 0156+2410     			& 0.013030  	& 54.9     	& -18.84     & 15.33     & 20.77  	& 0.49  & 5.3     & 1.6\\
UCM 1428+2727$^{\mathsf{d}}$  		& 0.014890 	& 62.6     	& -19.08     & 14.78     & 20.36  	& 0.24  & 5.2     & 1.7\\
UCM 1431+2854$^{\mathsf{d}}$     	& 0.031000 	& 127.9    	& -19.99     & 15.76     & 20.70  	& 0.52  & 4.4     & 2.7\\
UCM 1648+2855$^{\mathsf{d}}$     	& 0.032969 	& 134.6    	& -20.32     & 15.69     & 20.32  	& 0.26  & 3.9     & 2.4\\
UCM 2250+2427     			& 0.042149 	& 171.6    	& -21.59     & 15.40     & 20.23  	& 0.40  & 3.8     & 3.1\\
UCM 2258+1920    	 		& 0.021592 	& 90.1    	& -19.36     & 15.79     & 20.84  	& 0.21  & 4.4     & 2.0\\
UCM 2317+2356     			& 0.031962 	& 131.7  	& -21.91     & 14.16     & 20.90  	& 0.53  & 10.2    & 6.4\\
UCM 2327+2515n    			& 0.020600 	& 86.0       	& -19.23     & 15.79     & 19.63  	& 0.24  & 2.3     & 1.0\\
UCM 2327+2515s    			& 0.019130 	& 80.0       	& -19.06     & 15.80     & 20.30	& 0.43  & 3.2     & 1.3\\
SDSS 1134+1539				& 0.017335 	& 71.7 		& -19.91     & 14.47	 & 20.71	& 0.50 	& 7.3 	  & 2.5\\
SDSS 1507+5511				& 0.011231 	& 46.8 		& -19.23     & 14.19	 & 20.38	& 0.45 	& 7.1 	  & 1.6\\
SDSS 1605+4120				& 0.006640 	& 27.8 		& -18.65     & 13.63	 & 19.00	& 0.30 	& 4.8 	  & 0.7\\
SDSS 1652+6306				& 0.010376 	& 43.3 		& -18.48     & 14.78	 & 20.11	& 0.47	& 6.2 	  & 1.3\\
SDSS 1703+6127a				& 0.019690 	& 81.2 		& -19.62     & 14.74	 & 19.68	& 0.36 	& 4.0 	  & 1.6 \\
SDSS 1703+6127b				& 0.019760 	& 81.5 		& -18.40     & 16.20 	 & 21.13 	& 0.40 	& 4.0 	  & 1.5 \\
SDSS 1710+2138				& 0.014757 	& 61.2 		& -18.80     & 15.21	 & 20.69	& 0.53 	& 5.1 	  & 1.5\\
SDSS 2327-0923				& 0.016044 	& 66.5 		& -20.20     & 14.01	 & 20.25	& 0.58 	& 7.3 	  & 2.4\\
\hline
Mean$^{\mathsf{f}}$				& 0.0189 		& 78 		& -19.70 	     & 14.76	& 20.30          & 0.40              & 6.0             & 2.16\\
Standard deviation$^{\mathsf{f}}$ 	& $\pm0.0019$	& $\pm8$ 	& $\pm0.20$ & $\pm0.20$& $\pm0.11$ & $\pm0.02$   & $\pm0.6$ & $\pm0.25$\\
\hline
\multicolumn{9}{l}{$^{\mathsf{a}}$ Throughout the paper these galaxies are referred to by a truncation of the name listed here.} \\
\multicolumn{9}{l}{~~Typical uncertainties for the properties listed are, according to the SDSS and UCM catalogs,} \\
\multicolumn{9}{l}{~~$\pm$0.000025 for $z$, $\pm$0.1~Mpc for $d_{a}$, $\pm$0.10~mag for $M$($B$), $\pm$0.05~mag for $m$($b$),  $\pm$0.10~$B$-mag~arcsec$^{-2}$ for $SB_{e}$,}\\ 
\multicolumn{9}{l}{~~$\pm$0.07~dex for $B-V$, and $\pm$0.3~arcsec and $\pm$0.2~kpc for $R_e$.}\\
\multicolumn{9}{l}{$^{\mathsf{b}}$ Nasa Extragalactic Database; $^{\mathsf{c}}$ SDSS galaxies, NGC 6052, and NGC 469 properties from the SDSS catalog;}\\
\multicolumn{9}{l}{~~UCM galaxies, and NGC 7673 properties from the UCM catalog; NGC 7714 from \citet{garland05}}\\
\multicolumn{9}{l}{$^{\mathsf{d}}$ Also in the SDSS catalog; $^{\mathsf{e}}$ Also in the UCM catalog; $^{\mathsf{f}}$ Mean and standard deviation of the sample.}\\

\end{tabular}
\label{table:properties}
\end{table*}

\subsection{Observations}\label{observations}

Objects from our sample were observed using the PMAS spectrograph \citep{roth05} in the PPAK mode \citep{kelz06} at the 3.5-m telescope in CAHA\footnote{Based on the observations collected at the German-Spanish Astronomical Center, Calar Alto, jointly operated by the Max-Planck-Institut f\"ur Astronomie, Heidelberg, and the Instituto de Astrof\'isica de Andaluc\'ia (IAA/CSIC)}. PPAK is an integral field unit (IFU) consisting of 331 science fibers with a diameter of 2.7~arcsec each, covering a hexagonal FOV of $74\times65$~arcsec$^2$. Furthermore, there are 15 calibration fibers, which are always illuminated by a ThAr lamp and used to align images on the charge coupled device (CCD) camera; and 36 sky fibers located 80 arcsec away from the center of the hexagonal FOV. The PPAK bundle has gaps between each fiber and its neighbors. However, it is possible to fill these gaps by repeated observations of the same source with small pointing offsets; this is known as dithering.

PPAK observations of the 22 targets in our sample were made during three observing runs between 2005 and 2006 (Table~\ref{table:observations}). The first observing run (Run 47) took place during the nights of August 8 to 14, 2005. The second observing run (Run 56) took place during the nights of April 17 to 23, 2006. Finally, the third observing run (Run 64) took place during the nights of July 28 to August 3, 2006\footnote{Observing runs are referred to by their official CAHA names.}. Table ~\ref{table:observations} shows the coordinates for each galaxy, when and with what instrument configuration they were observed, and notes on the observing procedure when non-standard.

\begin{table*}
\centering
\caption{Sample Observations and Exposure Times per Dithering Position$^{\mathsf{a}}$}
\vspace{0.5cm}
\begin{tabular}{lcccccccl}
\hline
Name  			& V300  &	{}  & {}   &  {}  & V1200 	& {} & & {}  \\
	{}	& Run$^{\mathsf{b}}$ &	Exposure Times	&	{}	&	{}	& Run$^{\mathsf{b}}$ & Exposure Times	&	{}	&	{} \\
	{}	& {} &	d1 ($s$)	&	d2 ($s$)	&	d3 ($s$)	& {} & d1 ($s$)	&	d2 ($s$)	&	d3 ($s$)\\

\hline	
NGC 7673         & 47 & 990 & 990 & 990 & 47 & 2700 & 2700 & 2700\\	
NGC 7714         & 47 & 990 & 990 & 990 & 47 & 2700 & 2700 & 2700\\
NGC 6052         & 47 & 990 & 990 & 990 & 47 & 1800 & 1800 & 2700\\
NGC 469           & 47 & 990 & 990 & 990 & 47 & 2700 & 2700 & 2700\\
UCM 0000         & 64 & 990 & 990 & 990 & 64 & 7200 & \ldots & \ldots \\
UCM 0156         & 47 & 990 & \ldots & \ldots & 47 & 2700 & \ldots & \ldots \\
UCM 1428         & \ldots & \ldots & \ldots & \ldots & 56 & 3600 & 3600 & 3600\\
UCM 1431         & 47 & 990 & 990 & 990 & 47 & 2700 & 2700 & \ldots \\
UCM 1648         & 64 & 990 & 990 & 990 & 56 & 3600 & 3600 & 3600\\
UCM 2250         & 64 & 990 & 990 & 990 & 64 & 10800 & \ldots & \ldots \\
UCM 2258         & 64 & 990 & 990 & 990 & \ldots & \ldots & \ldots & \ldots \\
UCM 2317         & 47 & 990 & 990 & 990 & 47 & 2700 & 2700 & 2700\\
UCM 2327n       & 47 & 990 & 990 & 990 & 64 & 10800 & \ldots & \ldots \\
UCM 2327s       & 47 & 990 & 990 & 990 & 64 & 10800 & \ldots & \ldots \\
SDSS 1134        & \ldots & \ldots & \ldots & \ldots & 56 & 3600 & 3600 & 3600\\
SDSS 1507        & \ldots & \ldots & \ldots & \ldots & 56 & 3600 & 3600 & \ldots \\
SDSS 1605        & 47 & 990 & 990 & 990 & 64 & 10800 & \ldots & \ldots \\
SDSS 1652        & \ldots & \ldots & \ldots & \ldots & 64 & 10800 & \ldots & \ldots \\
SDSS 1703a      & 64 & 990 & 990 & 990 & 64 & 10800 & \ldots & \ldots \\
SDSS 1703b      & 64 & 990 & 990 & 990 & 64 & 10800 & \ldots & \ldots \\
SDSS 1710        & 64 & 990 & 990 & 990 & 47 & 2700 & 2700 & 2700\\
SDSS 2327        & 64 & 990 & 990 & 990 & 64 & 7200 & \ldots & \ldots \\
\hline
\multicolumn{9}{l}{$^{\mathsf{a}}$ Total exposure times are listed for each instrumental configuration and dithering position.}\\
\multicolumn{9}{l}{~~d1, d2, and d3 are dithering one, two, and three respectively.}\\
\multicolumn{9}{l}{~~Offsets, with respect to the coordinates of the objects, between dithering positions are}\\
\multicolumn{9}{l}{~~d1:(0.00, 0.00), d2: (+1.56, +0.78), and d3: (+1.56, -0.78).}\\
\multicolumn{9}{l}{$^{\mathsf{b}}$ Run 47: 08.08.05 to 08.14.05; Run 56: 04.17.06 to 04.23.06; Run 64: 07.28.06 to 08.03.06}\\ 
\end{tabular}
\label{table:observations}
\end{table*}

Two different configurations were used. First, a 300~lines~mm$^{-1}$ grating (V300) centered at 5316~{\AA} was used. This low resolution configuration provided a nominal spectral resolution of 10.7~{\AA} FWHM ($\sigma\sim255$~km~s$^{-1}$ at 5316~{\AA}) covering the spectral range from 3600 to 7000~{\AA}, including H$\beta$ and H$\alpha$. In some cases, three different dithering positions (i.e., d1, d2, and d3) were observed to fill the gaps between each fiber and its nearest neighbors. Three exposures, each 330~s, were taken for a total exposure time of 990~s per dithering position. 

Second, a 1200~lines~mm$^{-1}$ grating (V1200) centered at 5040~{\AA} was used. This high resolution configuration provided a nominal spectral resolution of 2.78~{\AA} FWHM ($\sigma\sim70$~km~s$^{-1}$ at 5040~{\AA}), covering the spectral range from 4669 to 5400~{\AA}, including H$\beta$ and [OIII]$\lambda$5007.

For a detailed summary of our observations, including telescope pointings and total exposure times for each galaxy, instrumental configuration, and dithering position, see Table~\ref{table:observations}

\subsection{Data Reduction}\label{reduction}

The raw data of fiber-fed spectrographs consist of a collection of spectra distributed along a certain axis of a two dimensional CCD frame. Each of the 382 spectra is dispersed along the x-axis. For each wavelength, each of the 382 spectra (i.e., science, sky, and calibration spectra) is also spread along the y-axis following a characteristic profile of finite width.

The data were reduced mainly using R3D\footnote{http://www.caha.es/sanchez/r3d/} and E3D\footnote{http://www.aip.de/Euro3D/E3D/} \citep{sanchez06}, while IRAF\footnote{IRAF is distributed by the National Optical Astronomy Observatory, which is operated by the Association of Universities for Research in Astronomy (AURA) under cooperative agreement with the National Science Foundation.}, and our
own custom software were also used. All the images were bias-subtracted, flat-fielded, and cosmic-ray cleaned. The 331 science spectra per dithering position were then properly extracted, distortion-corrected, wavelength-calibrated, sky-subtracted, and flux-calibrated, as summarized below.

\textbf{Bias-subtraction.} Bias images were taken, averaged, and subtracted to correct our data images. IRAF was used for these and most algorithmic tasks throughout our data reduction process.

\textbf{Flat-fielding.} Flat-field images were obtained by illuminating the dome with a continuum lamp.  Due to flexures of the instrument, flat-field images were taken before every object and calibration lamp exposure. 

\textbf{Cosmic-ray cleaning.} By combining three different exposures per dithering we were able to remove cosmic rays from our images. This was done using a sigma clipping algorithm. When fewer than three exposures were available per field IRAF task L.A. Cosmic \citep{dokkum01} was used.

The fundamental steps in IFS data reduction (i.e., extraction, distortion, and wavelength calibration) were done by using R3D \citep{sanchez06}.

\textbf{Extraction.} The location of each spectrum was found on the detector for each pixel along the dispersion axis by considering a Gaussian profile in order to proceed to its extraction. Homogeneously illuminated flat-field images were used to find and trace the position of the spectrum in the raw data. Cross-talk between fibers makes this step crucial. We used an algorithm that minimizes the contamination due to this effect \citep{sanchez06}.

\textbf{Distortion-correction.} For grating spectrographs, the entrance slit is distorted and imaged as a curve onto the CCD. In order to correct our data for this effect, the intensity peak of a set of selected emission lines from our calibration lamp exposures is traced, and a distortion correction is determined to re-center all the lines to a common reference. These distortion corrections are subsequently applied to the science exposures. Nonetheless, we needed first to align our calibration arc and science exposures by matching the intensity peaks of a set of selected ThAr emission lines from the 15 calibration fibers.

\textbf{Wavelength-calibration.} The wavelength coordinate system was determined by identifying the wavelengths of the arc emission lines.

The V300 wavelength calibration was performed using a He lamp with up to 17 lines within the considered spectral range. The rms of the best fit polynomial ($n=3$) was 0.12~{\AA}. A different arc was obtained every single night; these values were consistent up to the second significant figure. A further analysis to evaluate the accuracy of our calibration including sky lines gave a final uncertainty in our calibration of 0.2~{\AA} (10~km~s$^{-1}$ at 6000~{\AA}). For this analysis, five high S/N sky lines from the wavelength-calibrated spectra covering the entire V300 spectral range were fitted by single Gaussian profiles for each fiber. The centroids of these lines were then compared to those of the CAHA sky atlas \citep{sanchez07}. The uncertainty of our measurements was given by the standard deviation of the resulting residuals, which was consistent within the entire V300 spectral range (i.e., for each sky line independently).

On the other hand, the V1200 wavelength calibration was performed using both He and Cs lamps because the He lamp lacked emission lines bluer than 5015~{\AA}. During the first run, the Cs lamp was observed separately and both lamps were added to increase the spectral coverage. Furthermore, weak contamination lines from the ThAr lamp used to illuminate the calibration fibers were also used for this purpose. Despite our efforts to produce an arc frame by combining different arcs, the reduced, poor quality of the lines and the inability to produce this calibration frame at the same telescope position as the actual observations, the rms of the best fit polynomial ($n=3$) translated into a higher uncertainty of at least 0.25~{\AA} (15~km~s$^{-1}$ at 5000~{\AA}). 

A better wavelength calibration was possible during the last two runs. It was performed using both He and Cs lamps simultaneously with up to 12 lines within the considered spectral range. Although observing both lines at the same time improved our wavelength calibration, this was still far from what we would have expected due to the inappropriate set of arcs available. The rms of the best fit polynomial ($n=3$) was 0.06~{\AA} (4~km~s$^{-1}$ at 5000~{\AA}). A different arc was obtained every single night; these values were consistent up to the second significant figure. 

The lack of sky lines in the V1200 spectral range did not allow us to carry out a similar analysis to the one carried out for the V300. However, our data can be wavelength-calibrated using the rest-frame positions of H$\beta$, [OIII]$\lambda$4959, and [OIII]$\lambda$5007. We wavelength-calibrated each of the fibers independently. When this was done, the rms for this particular spectral range was 0.13~{\AA} (8~km~s$^{-1}$ at 5000~{\AA}).

Taking all this into account, and to be consistent, we decided to use V1200 data only for velocity width measurements, and velocity measurements when V300 data were not available. (Although velocity measurements were also made for all the available V1200 data to check that they were in agreement with those from V300.) V300 data were used for velocity measurements because they allowed better sampling of the galaxies by means of the dithering technique applied in almost all V300 observations.

\textbf{Sky-subtraction.} The 36 sky spectra provided by the 36 sky fibers located 80~arcsec away from the center of the hexagonal FOV were averaged and subtracted from the object spectra.

\textbf{Flux-calibration.} The PPAK fiber bundle does not cover the entire FOV because it has gaps between each fiber and its nearest neighbors. However, our dithering technique allows us to cover the entire FOV with three offset telescope pointings. Therefore, it is possible to determine a relative calibration for each dithering using a standard star observed with PPAK, and re-calibrate the spectra using broad-band photometry \citep{castillo10}.

The data reduction process provided 331 fully reduced science spectra per dithering position and per galaxy. This is 2,979 spectra per galaxy, and 65,583 spectra in total.

\subsection{Basic Measurements}\label{measurements}

Emission lines were fitted by single Gaussian profiles both in the V300 (H$\alpha$) and V1200 ([OIII]$\lambda$5007) configurations. A minimum S/N of 13 in the flux detection was required for the measurements to be used. According to simulations we carried out, below this threshold, the uncertainty associated with a poor S/N dominates over the wavelength calibration uncertainty as the main source of error. These simulations were done by adding noise to an emission line with high S/N. By the time the S/N ratio dropped to 13 it was the poor S/N which dominated the uncertainty associated with the position of the centroid when the emission line was fit by a Gaussian profile. 
 
\begin{figure*}
  \begin{center}
\begin{tabular}{ccc}

      \includegraphics[width=5.cm]{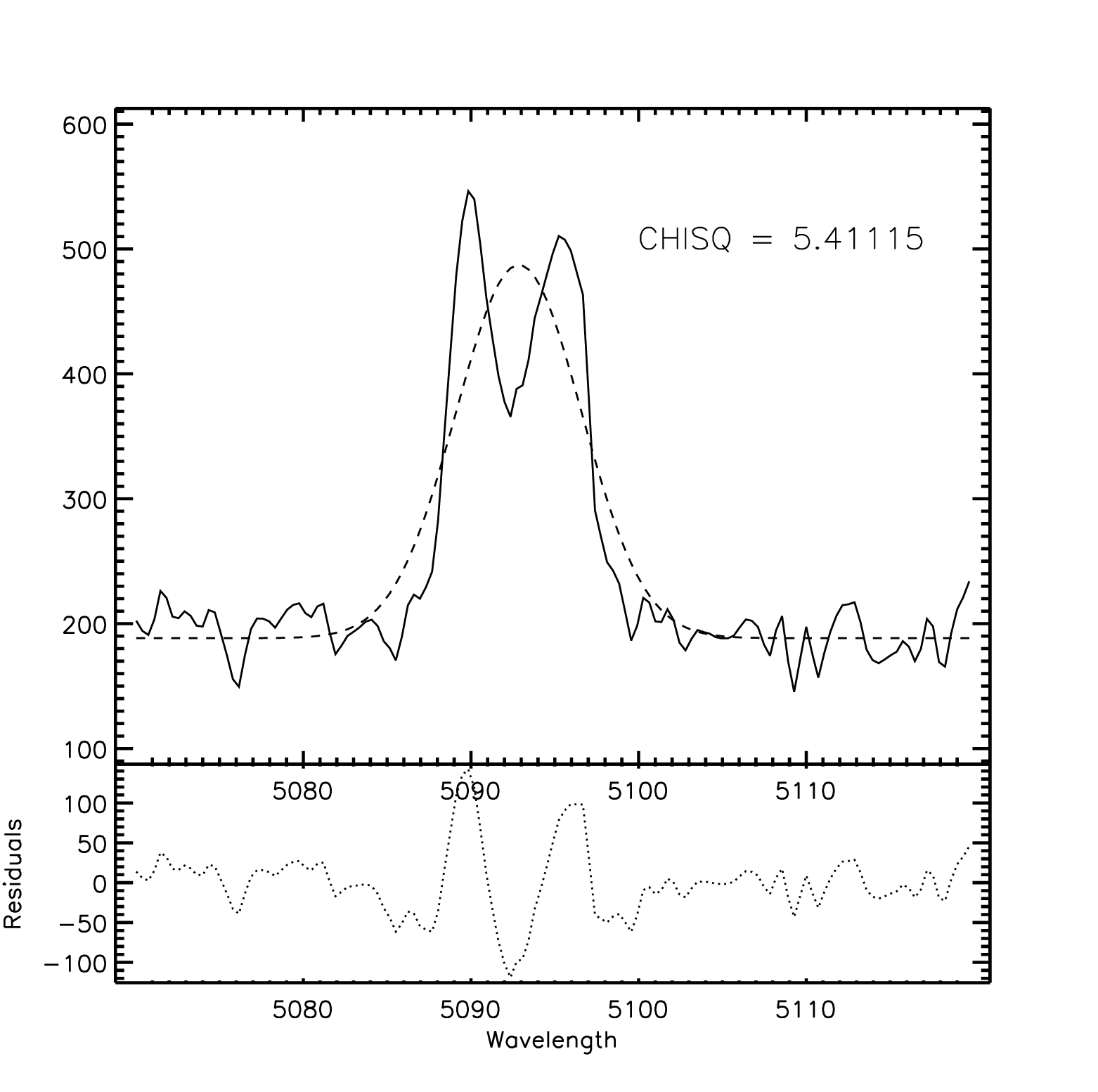} &
      \includegraphics[width=5.cm]{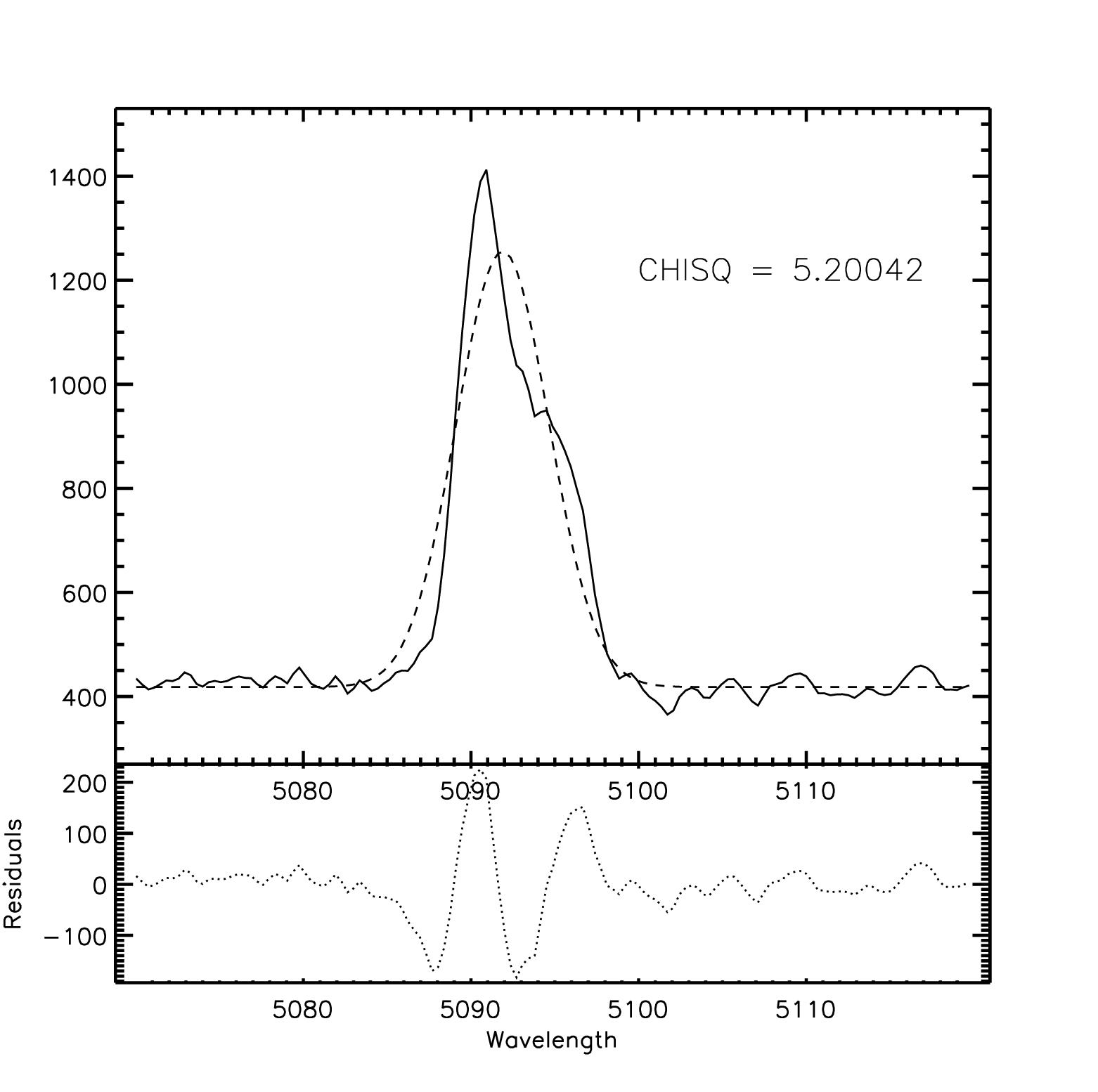} &
      \includegraphics[width=5.cm]{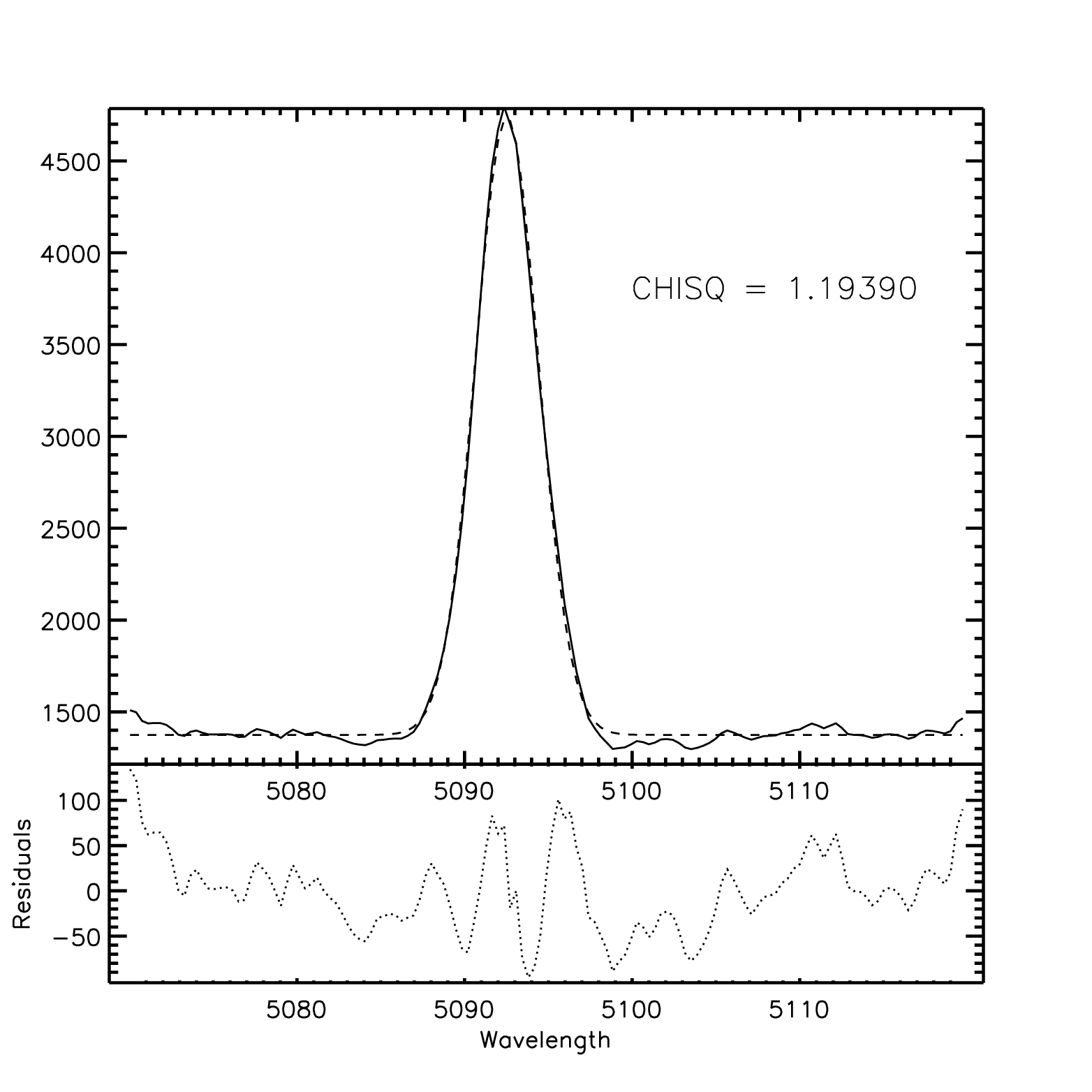} \\ 

      \includegraphics[width=5.cm]{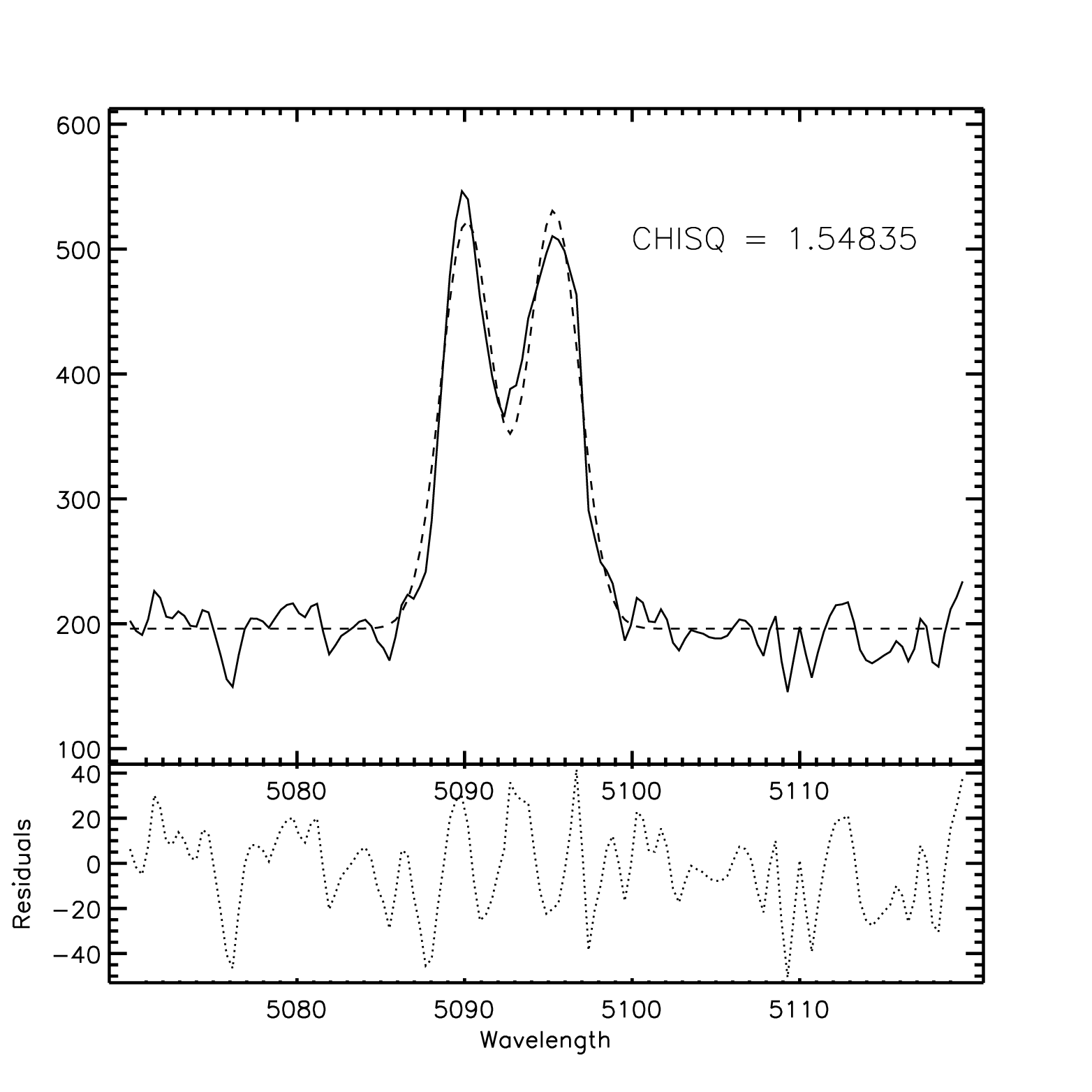} &    
      \includegraphics[width=5.cm]{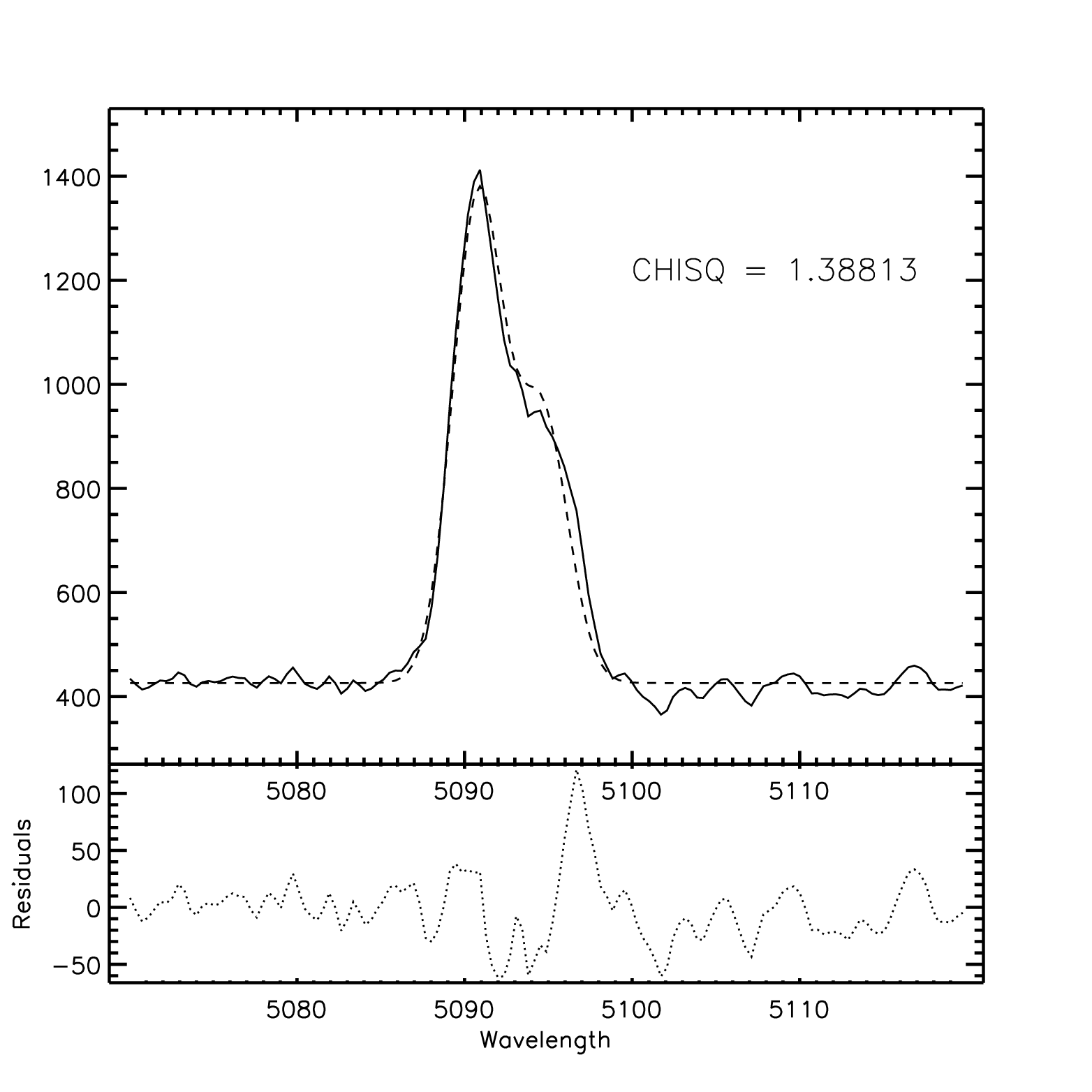} &
      \includegraphics[width=5.cm]{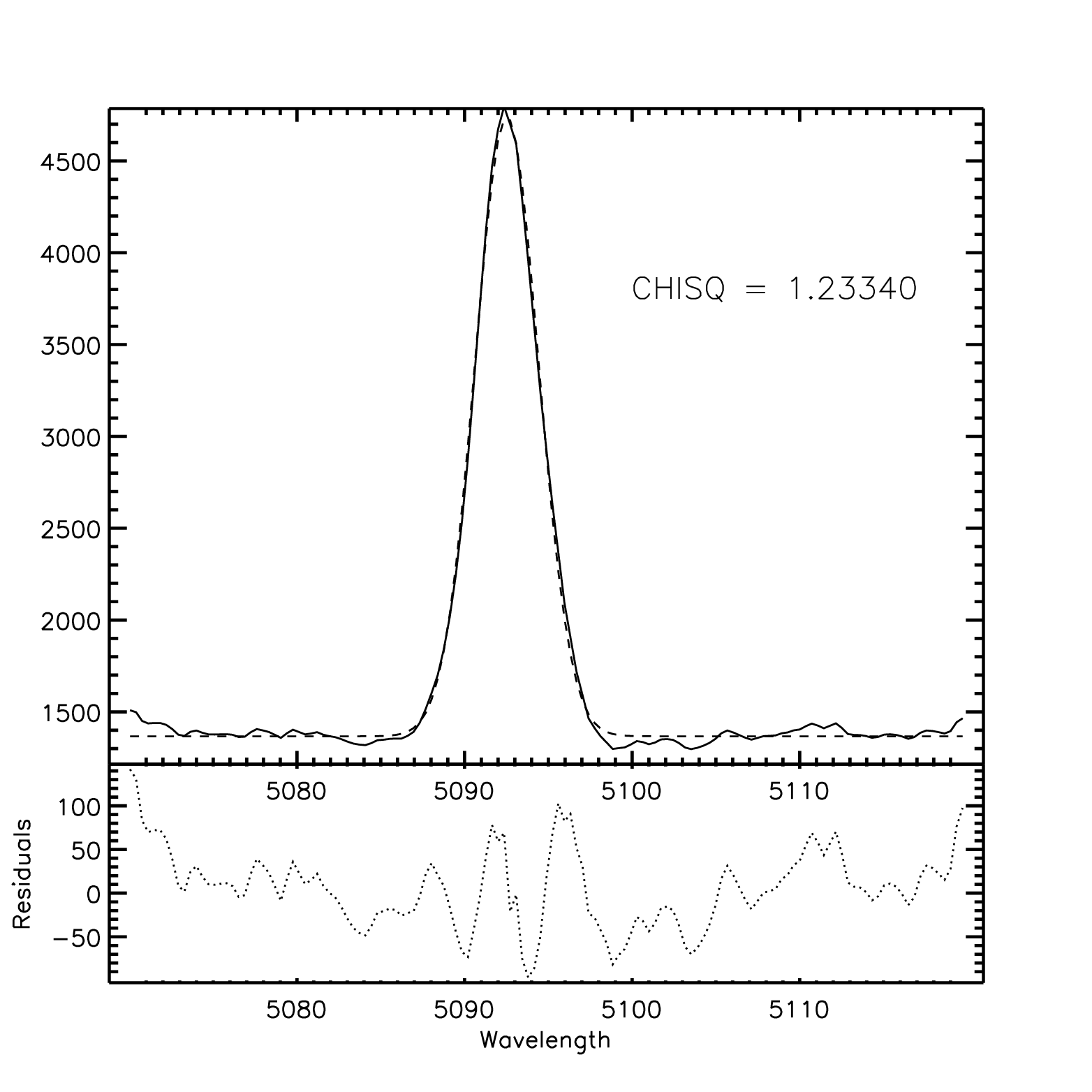} \\ 

\end{tabular}      
    \caption{\emph{Left column:} Example of two blended emission lines which are easy to identify. Single (\emph{top}) and double (\emph{bottom}) Gaussian profile fits are shown (dashed line). \emph{Middle column:} Example of two blended emission lines. Single (\emph{top}) and double (\emph{bottom}) Gaussian profile fits are shown (dashed line). \emph{Right column:} Two examples of single Gaussian profile fits (dashed line). The $\chi^2$ (CHISQ) of each fit is shown, as well as the residuals for all the fits in the bottom box of each plot. The flux scale (y-axis) is arbitrary and the wavelength (x-axis) is in {\AA}.}

    \label{fig:two_gauss}
  \end{center}
\end{figure*}

For the data taken in the V1200 configuration, the presence of double emission line components was addressed by means of the calculation of the reduced $\chi^2$ (CHISQ) of single and double Gaussian profile fits.  After carefully investigating our data we decided to attempt a double Gaussian profile fit whenever the reduced $\chi^2$ of our single Gaussian profile fit was $\gg$1 (Figure~\ref{fig:two_gauss}). In order to guarantee the reliability of the reduced $\chi^2$ as an indicator of non-Gaussianity and to avoid contamination, we required a minimum S/N of 30 for double Gaussian profile fits to be used.

The velocity fields of the galaxies were created by measuring the centroids of the H$\alpha$ emission lines for each dithering position with the low resolution spectra. The measurements for each dithering position were then interpolated down to a spatial resolution of 1 arcsec pixel$^{-1}$ yielding a 60 $\times$ 60~pixel$^2$ square grid. Each of the original fibers was therefore sampled by approximately 3 $\times$ 3~pixel$^2$ (i.e., $\sim2.7\times 2.7$~arcsec$^2$). Oversampling the data beyond these values may result in the appearance of artifacts (S. S\'anchez, private communication). The maps for each dithering position were then registered and averaged pixel by pixel. Thus, both the resolution and the error associated with each measurement improved by a factor of $\sqrt{3}$. Thus, the final uncertainties are 6~km~s$^{-1}$ for V300, and 2 and 5~km~s$^{-1}$ for V1200 depending on the wavelength calibration. When only two dithering positions were available, the velocity map resulted from combining two sets of interpolated data, with a final improvement of a factor of $\sqrt{2}$. When only one dithering was available, the velocity map resulted from the direct interpolation of the measurements of the velocity for each of its fibers. Finally, for four galaxies, V1200 data ([OIII]$\lambda$5007 emission lines) were used to produce these maps since V300 data were not available.

In the outer areas of all galaxies, we spatially binned the data by co-adding fibers to achieve a minimum S/N of 13. Each of these new measurements was linked to the average position of the co-added fibers. 
Thus, the maps extend towards the lower surface brightness outskirts of the galaxy, gaining an area about 10\% larger per galaxy.
 
The velocity width maps of the galaxies were derived from the [OIII]$\lambda$5007 emission lines for each dithering position with the high resolution spectra. This line was selected as the one with the highest S/N within the V1200 spectral range. The instrumental broadening $(\sigma_{instrument}$), as measured in sky lines for each fiber, was subtracted in quadrature from the measured broadening ($\sigma_{measured}$) to find the intrinsic broadening of each measurement ($\sigma_{intrinsic}$). It is important to note that the average measured instrumental broadening (2.3~{\AA} FWHM) was considerably smaller than the nominal value listed in the instrument manual (2.78~{\AA} FWHM). The final map was produced as explained above except that no binning was performed in this case since it would have artificially broadened the emission lines. This translates, in general, to a smaller velocity width area. The observational strategy for the V1200 configuration is different for different galaxies, leading in some cases to longer exposures and higher S/N, which explains why the FOV is not always smaller (e.g., for UCM 2327 the area of the V1200 is larger than for the V300). Regarding the velocity width uncertainties, a similar analysis as the one described for the S/N threshold shows typical uncertainties of 10\% ($\sim$7~km~s$^{-1}$) for individual spectra with low S/N (i.e., $\sim$13), and down to $\sim$2\% ($\sim$1~km~s$^{-1}$) for high S/N spectra, such as those used for the measurements of the integrated velocity width of our galaxies.

\section{Data Analysis}\label{statistical}

Optical images from the SDSS\footnote{Based on observations obtained with the Sloan Digital Sky Survey and with the Apache Point Observatory 3.5 m telescope, which is owned and operated by the Astrophysical Research Consortium. False color images were retrieved from http://cas.sdss.org/astro/en/tools/chart/navi.asp} or the Digitized Sky Survey (DSS)\footnote{Based on data mining of the Digitized Sky Survey, developed and operated by the Catalogs and Surveys Branch of the Space Telescope Science Institute, Baltimore, Maryland. Images were retrieved through the NASA Extragalactic Database.}, velocity maps, and velocity width maps for our sample are shown in Figures 2 to 7. For each galaxy, the size and orientation of the FOV of the optical image are the same as those of the FOV of the velocity and velocity width maps. For UCM 2258 only, the velocity map is shown because high resolution data were not available. The results of these observations are summarized in Tables~\ref{table:statistics1} and~\ref{table:statistics2}.

Galaxies UCM 2327n and UCM 2327s were considered as one galaxy (UCM 2327). This pair is separated by approximately 0.2~Mpc (less than 10~arcsec), as seen in the FOV of PPAK.

\subsection{Kinematic Classification}\label{classification}

In order to classify the kinematic maps of distant starbusrts \citet{yang08} defined three different groups:

\begin{itemize}
\item{\bf Rotating disks (RD).} The velocity map shows an ordered gradient and the dynamical major axis is aligned with the morphological major axis. The velocity width map indicates a single peak close to the dynamic center.

\item{\bf Perturbed rotation (PR).} The kinematics show all the features of an RD, but the peak in the velocity width map is either absent or clearly shifted away from the dynamical center.

\item{\bf Complex kinematics (CK).} Neither the velocity map nor the velocity width map are compatible with regular disk rotation. The velocity maps are misaligned with the morphological major axis.  
\end{itemize}

\citet{yang08} found, in a sample of 63 galaxies, 32\% RDs, 25\% PRs, and 43\% CKs, with an uncertainty of 12\%, confirming that at $0.4<z<0.75$, only one-third of massive starbursts are kinematically relaxed. Using similar classification criteria, \citet{forster09} found a similar distribution for a sample of 63 galaxies at $1.3<z<2.6$.
However, the apparent size of these objects makes it difficult to fit their velocity and velocity width maps with models uniquely.

\begin{figure*}\label{atlas1}
  \begin{center}
\begin{tabular}{ccc}

      \includegraphics[width=4.5cm]{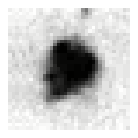}&
      \includegraphics[width=6.cm]{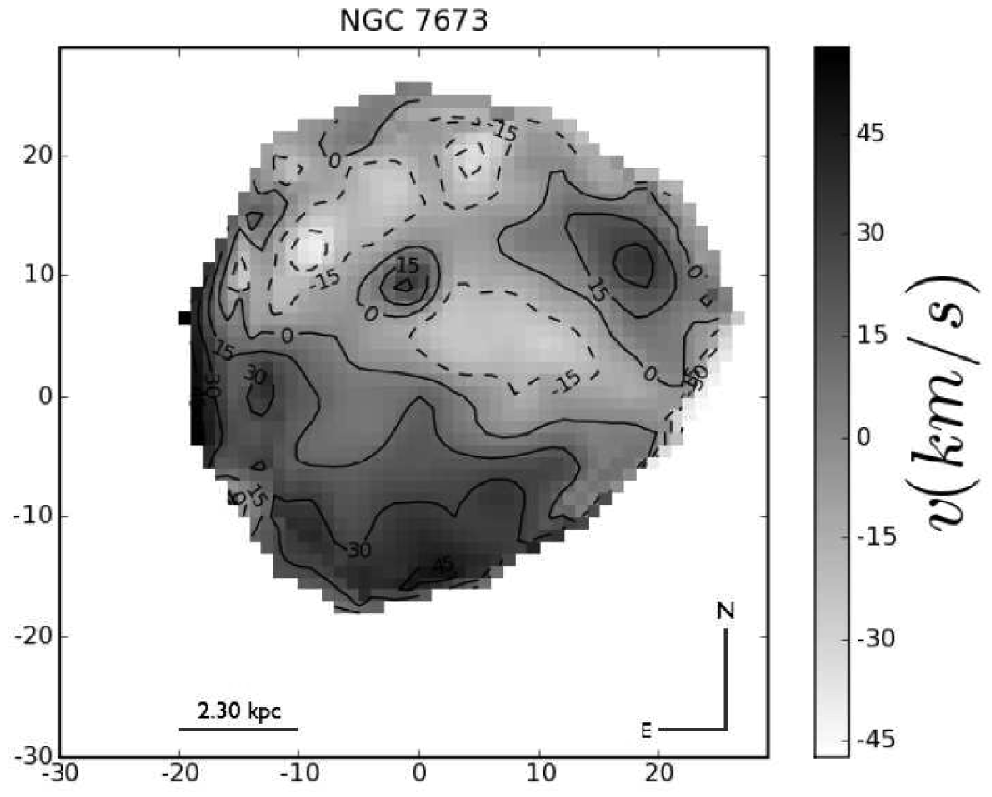}&  
      \includegraphics[width=6.cm]{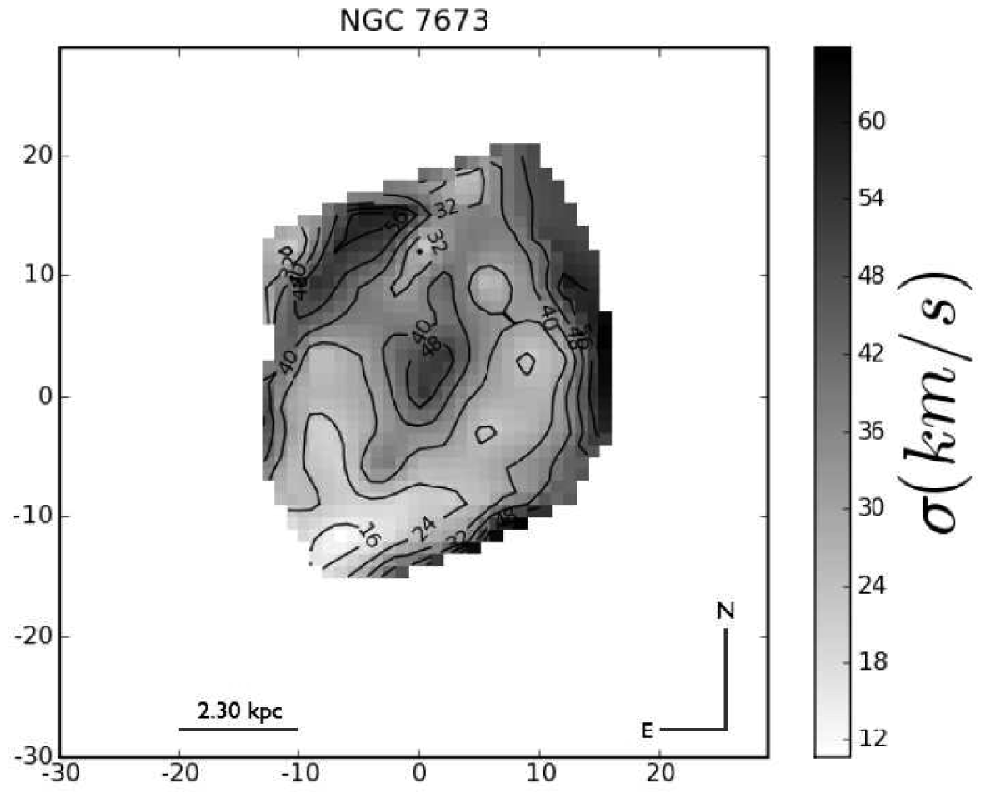}\\

      \includegraphics[width=4.5cm]{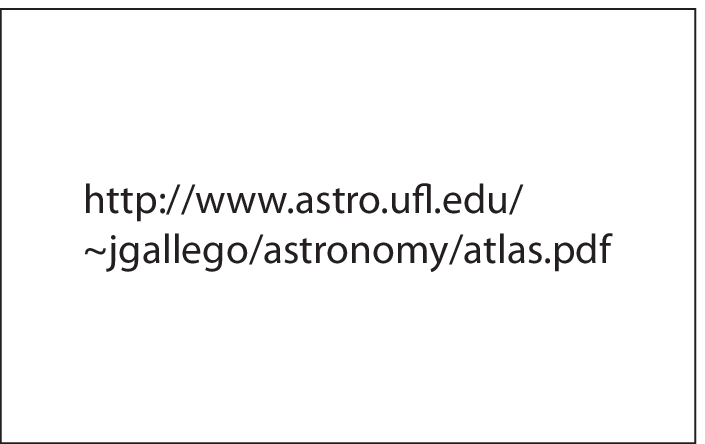}&  
      \includegraphics[width=6.1cm]{./blank.eps}&  
      \includegraphics[width=6.1cm]{./blank.eps}\\

      \includegraphics[width=4.5cm]{./blank.eps}&  
      \includegraphics[width=6.1cm]{./blank.eps}&  
      \includegraphics[width=6.1cm]{./blank.eps}\\

      \includegraphics[width=4.5cm]{./blank.eps}&  
      \includegraphics[width=6.1cm]{./blank.eps}&  
      \includegraphics[width=6.1cm]{./blank.eps}\\  

\end{tabular} 
\caption{For each galaxy the following images are shown: (\emph{left}) the SDSS or DSS images; (\emph{center}) the velocity map; and (\emph{right}) the velocity width map. The size and orientation of the FOV of the optical image are the same as those of the FOV of the velocity and velocity width maps. Axis units are arcseconds. North is up and east is to the left. Contours follow the velocity and velocity dispersion gradients. For the former, dashed contours correspond to negative velocities (i.e., with respect to the redshift of each galaxy), while solid contours correspond to positive velocities.}

\end{center}
\end{figure*}

\begin{figure*}
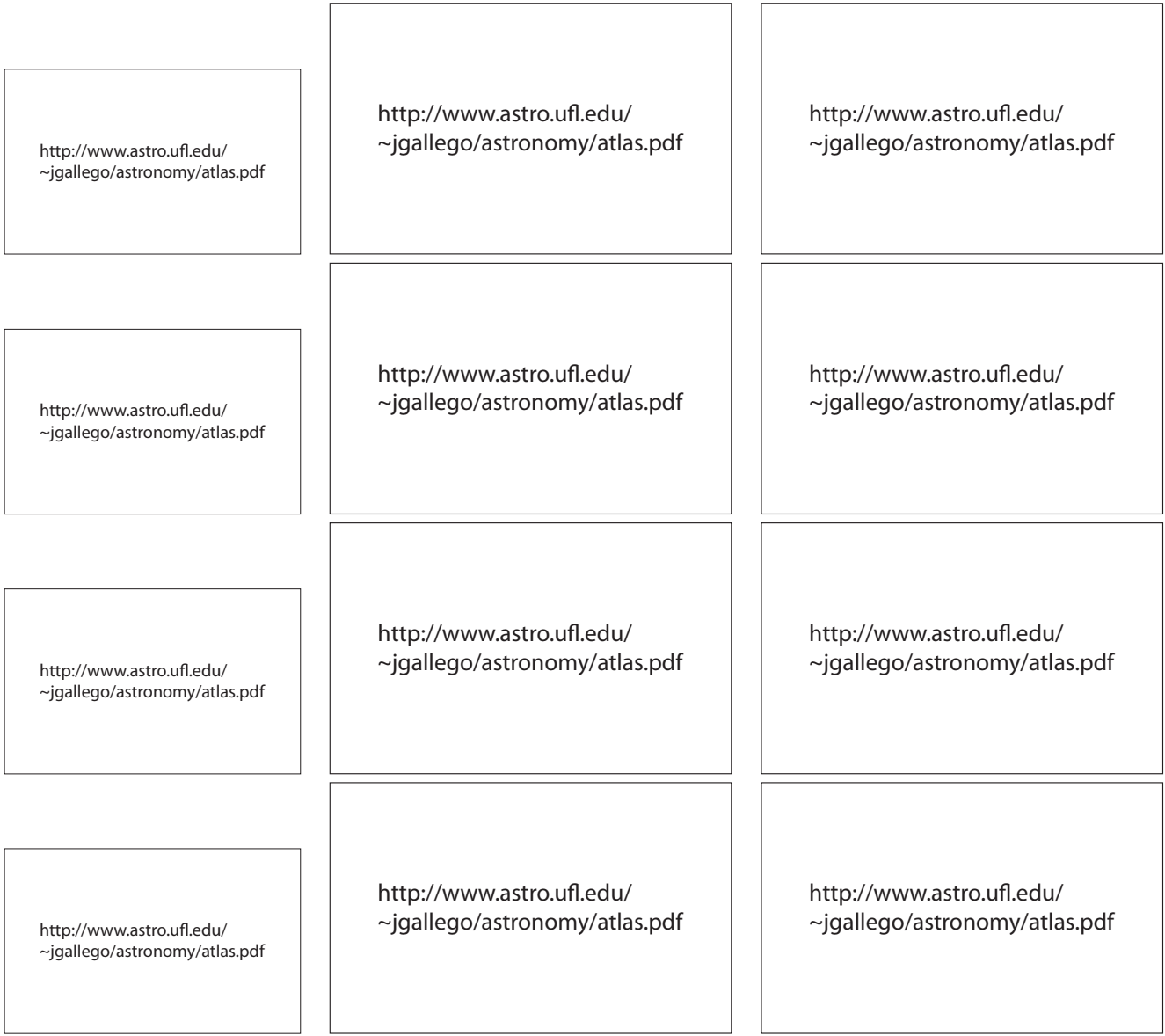
\label{atlas2}
  \begin{center}
\begin{tabular}{ccc}

      \includegraphics[width=4.5cm]{./blank.eps}& 
      \includegraphics[width=6.1cm]{./blank.eps}&  
      \includegraphics[width=6.1cm]{./blank.eps}\\

      \includegraphics[width=4.5cm]{./blank.eps}&  
      \includegraphics[width=6.1cm]{./blank.eps}&  
      \includegraphics[width=6.1cm]{./blank.eps}\\

      \includegraphics[width=4.5cm]{./blank.eps}&  
      \includegraphics[width=6.1cm]{./blank.eps}&  
      \includegraphics[width=6.1cm]{./blank.eps}\\  

      \includegraphics[width=4.5cm]{./blank.eps}&  
      \includegraphics[width=6.1cm]{./blank.eps}&  
      \includegraphics[width=6.1cm]{./blank.eps}\\  

\end{tabular} 
\caption{Same as Figure 2. The UCM 1428 offset is due to this galaxy accidentally being observed off center from the PPAK FOV. }
\end{center}

\end{figure*}

\begin{figure*}
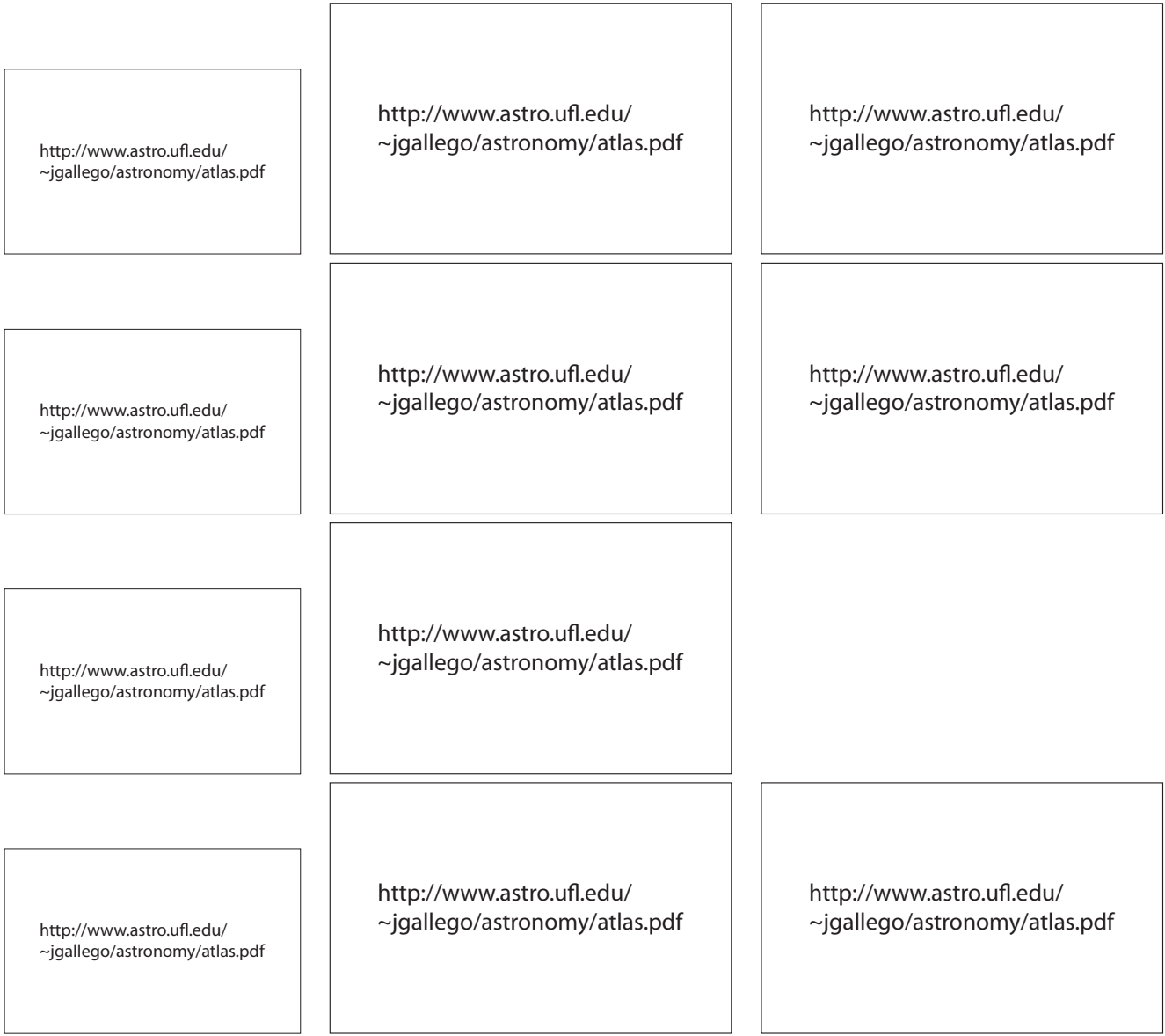
\label{fig:atlas3}
  \begin{center}
\begin{tabular}{ccc}

      \includegraphics[width=4.5cm]{./blank.eps}&  
      \includegraphics[width=6.1cm]{./blank.eps}&  
      \includegraphics[width=6.1cm]{./blank.eps}\\  

      \includegraphics[width=4.5cm]{./blank.eps}&  
      \includegraphics[width=6.1cm]{./blank.eps}&  
      \includegraphics[width=6.1cm]{./blank.eps}\\  

      \includegraphics[width=4.5cm]{./blank.eps}&  
      \includegraphics[width=6.1cm]{./blank.eps}&\\  

      \includegraphics[width=4.5cm]{./blank.eps}& 
      \includegraphics[width=6.1cm]{./blank.eps}&  
      \includegraphics[width=6.1cm]{./blank.eps}\\ 

\end{tabular} 
\caption{Same as Figure 2. For UCM 2258 only the velocity map is shown because high resolution data were not available.} 
\end{center}

\end{figure*} 

\begin{figure*}
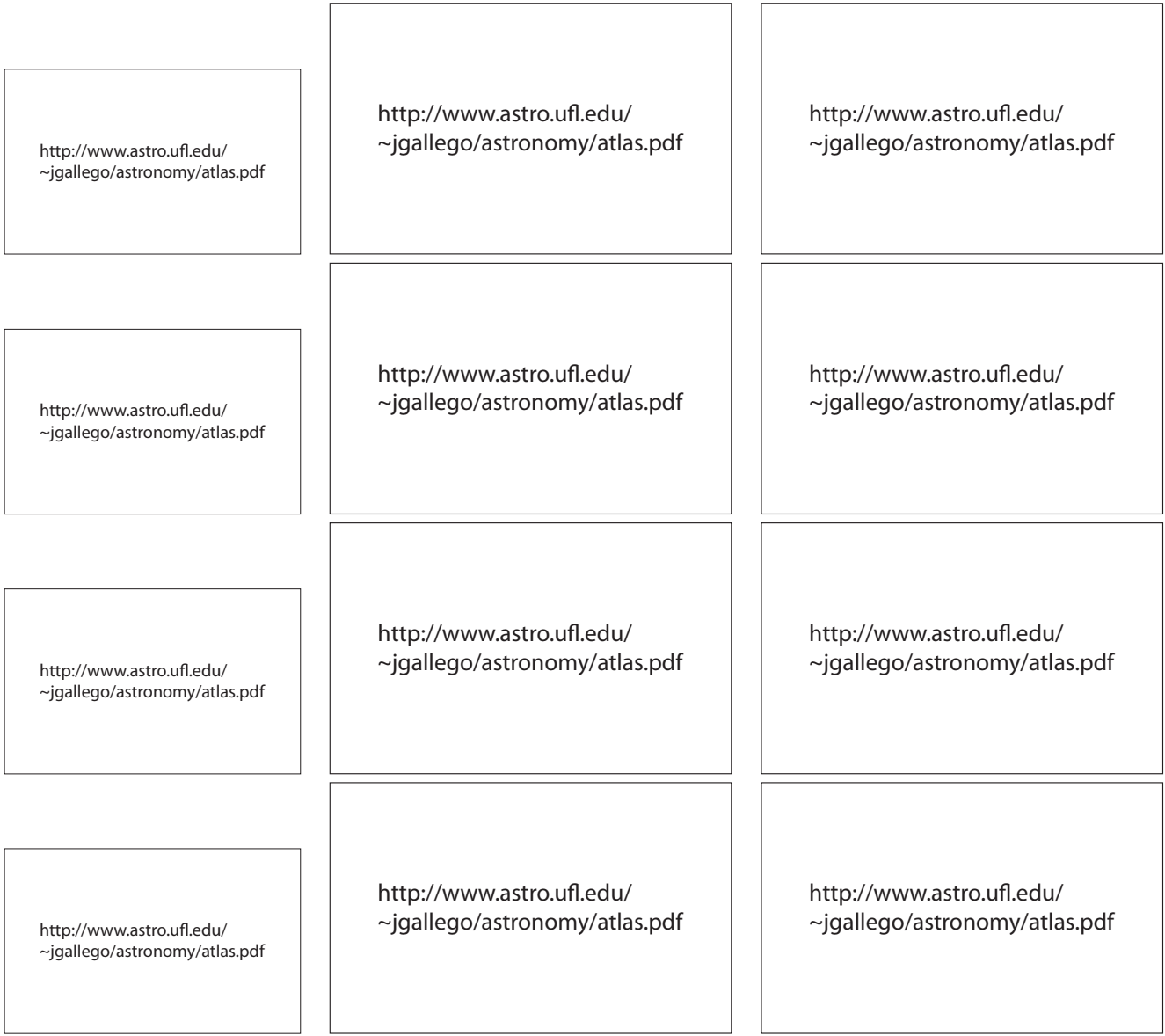
\label{fig:atlas4}
  \begin{center}
\begin{tabular}{ccc}

      \includegraphics[width=4.5cm]{./blank.eps}&  
      \includegraphics[width=6.1cm]{./blank.eps}&  
      \includegraphics[width=6.1cm]{./blank.eps}\\  

      \includegraphics[width=4.5cm]{./blank.eps}&  
      \includegraphics[width=6.1cm]{./blank.eps}&  
      \includegraphics[width=6.1cm]{./blank.eps}\\  

      \includegraphics[width=4.5cm]{./blank.eps}&  
      \includegraphics[width=6.1cm]{./blank.eps}&  
      \includegraphics[width=6.1cm]{./blank.eps}\\  

      \includegraphics[width=4.5cm]{./blank.eps}&  
      \includegraphics[width=6.1cm]{./blank.eps}&  
      \includegraphics[width=6.1cm]{./blank.eps}\\  

\end{tabular} 
\caption{Same as Figure 2. The SDSS 1134 offset is due to trying to detect a second smaller galaxy, visible in the SDSS image, within the FOV of PPAK. Both UCM 2327n and UCM 2327s fit within the FOV of PPAK and are shown as UCM 2327.}
\end{center}

\end{figure*}

\begin{figure*}\label{fig:atlas5}
  \begin{center}
\begin{tabular}{ccc}

      \includegraphics[width=4.5cm]{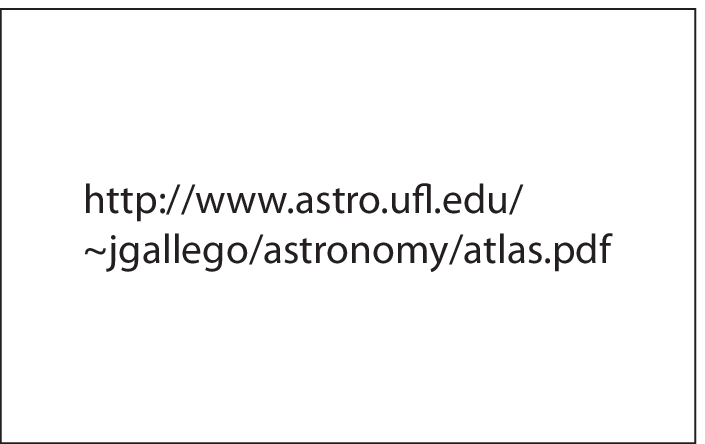}&  
      \includegraphics[width=6.1cm]{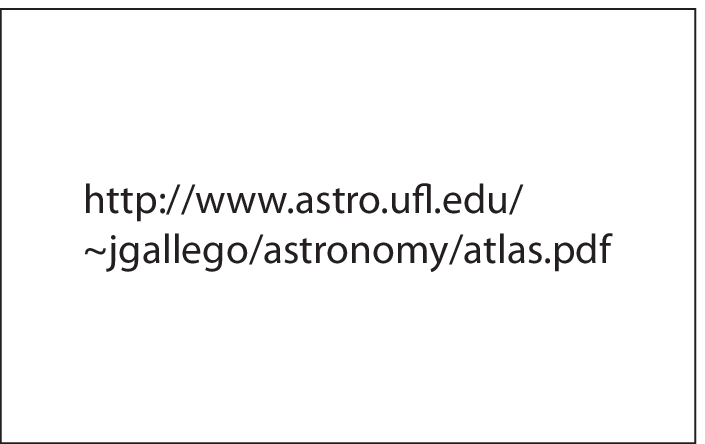}&  
      \includegraphics[width=6.1cm]{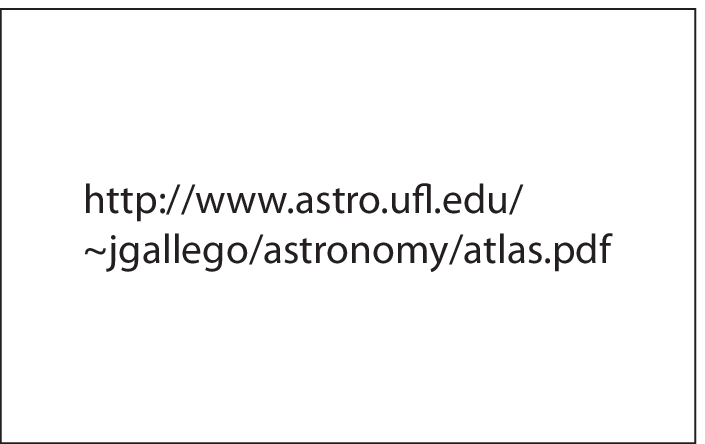}\\  

      \includegraphics[width=4.5cm]{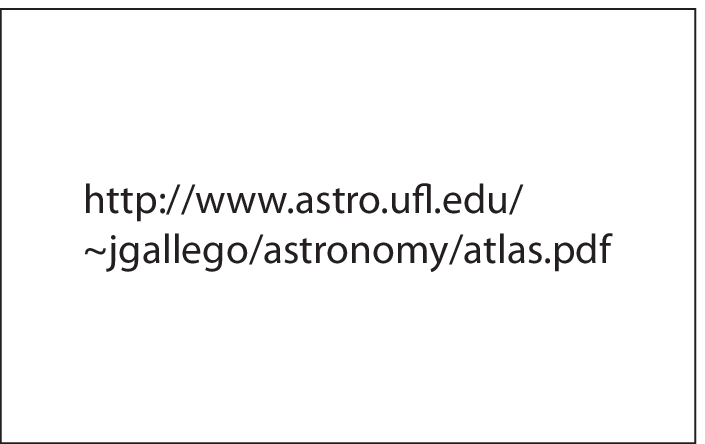}&  
      \includegraphics[width=6.1cm]{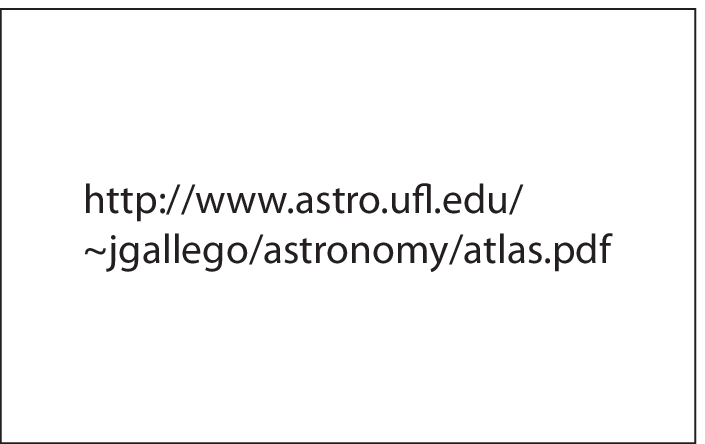}&  
      \includegraphics[width=6.1cm]{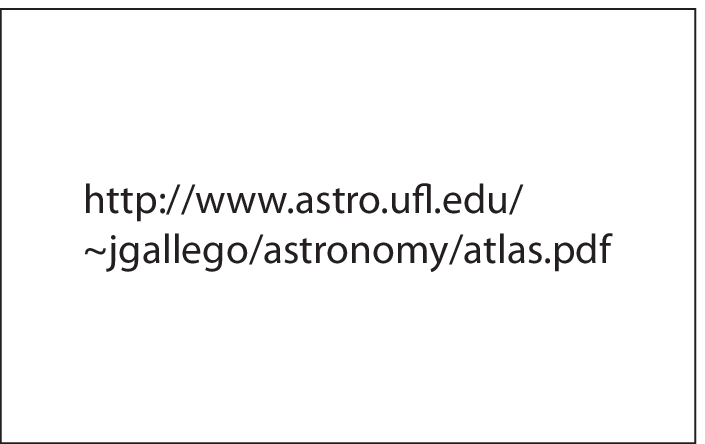}\\  

      \includegraphics[width=4.5cm]{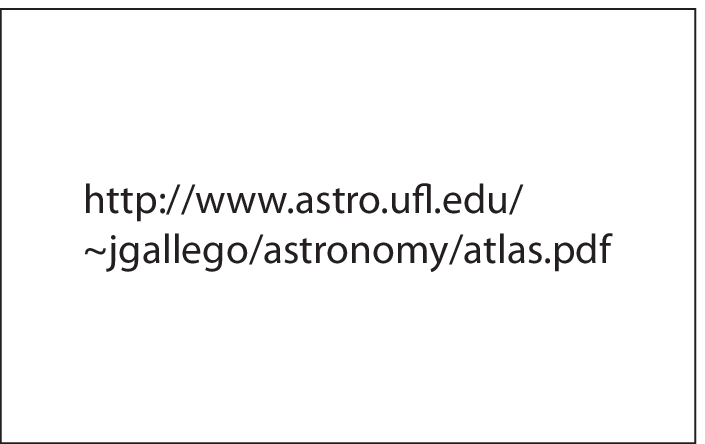}&  
      \includegraphics[width=6.1cm]{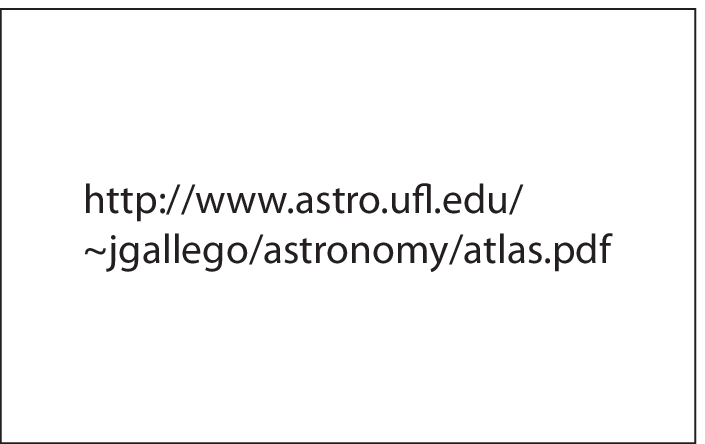}&  
      \includegraphics[width=6.1cm]{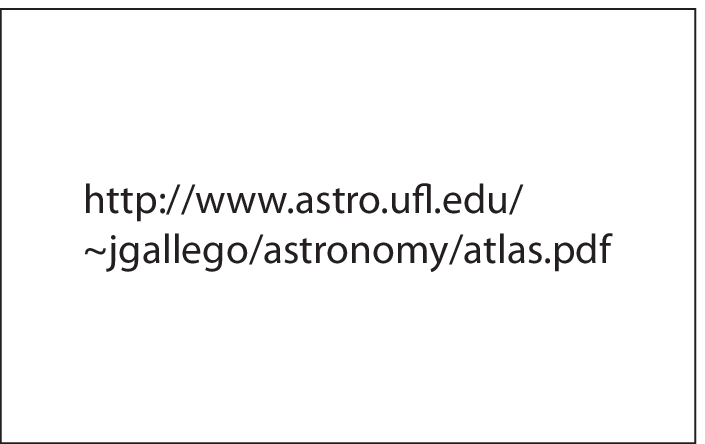}\\  

      \includegraphics[width=4.5cm]{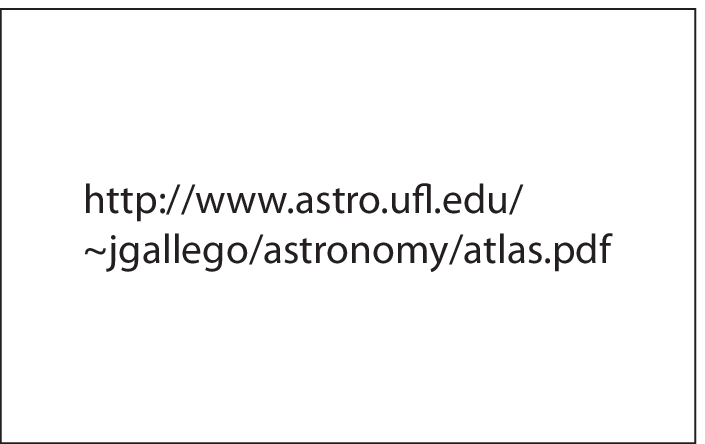} & 
      \includegraphics[width=6.1cm]{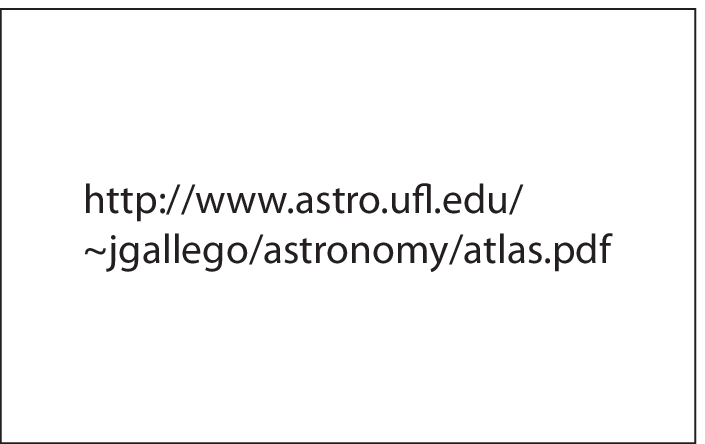} & 
      \includegraphics[width=6.1cm]{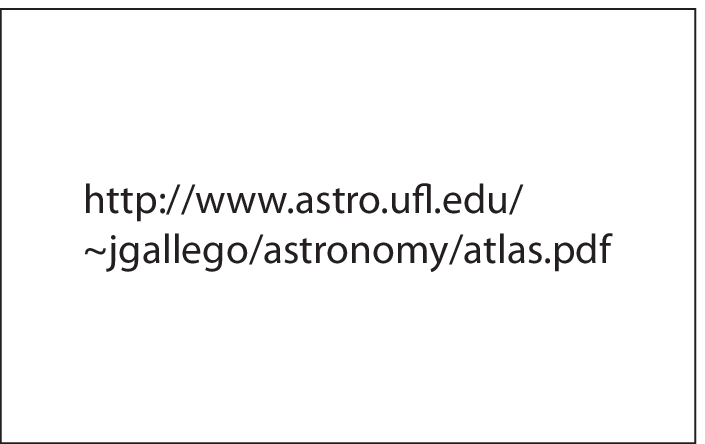} \\ 

\end{tabular} 
\caption{Same as Figure 2.}
\end{center}

\end{figure*}

\begin{figure*}\label{fig:atlas6}
  \begin{center}
\begin{tabular}{ccc}

      \includegraphics[width=4.5cm]{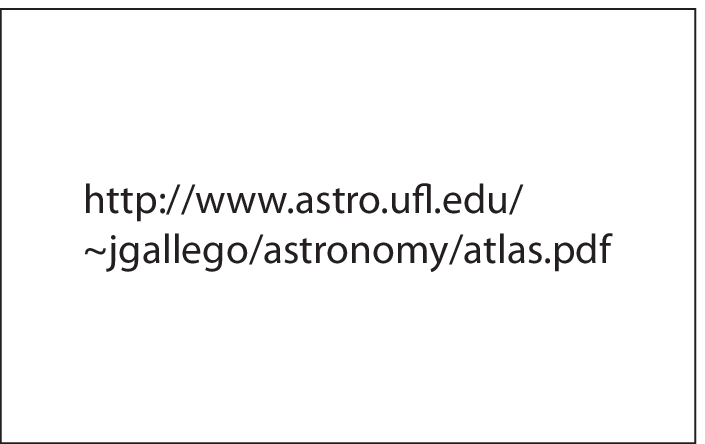}  &
      \includegraphics[width=6.1cm]{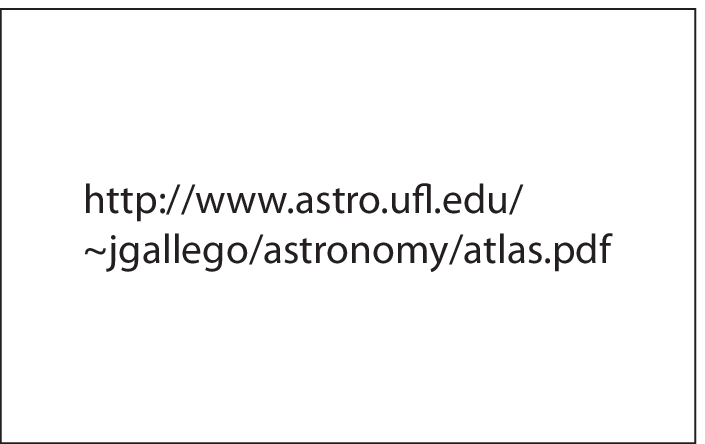}  &
      \includegraphics[width=6.1cm]{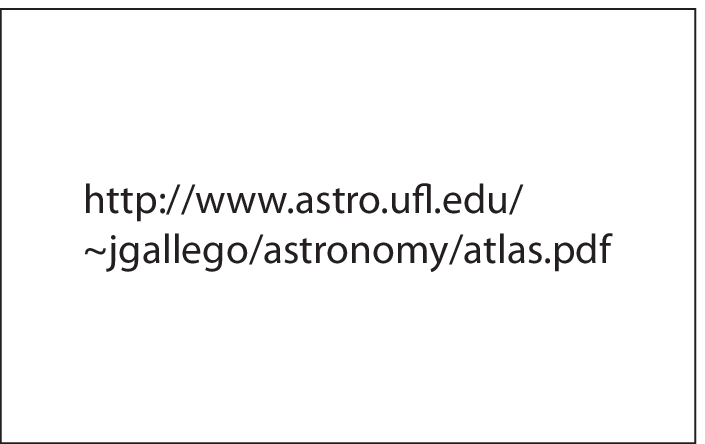}  \\

\end{tabular} 
\caption{Same as Figure 2.}
\end{center}

\end{figure*}

\begin{table*}
\centering
\caption{Kinematics and Morphology}
\vspace{0.5cm}
\begin{tabular}{lcccclll}
\hline
Name 	& $v_{rot}/\sin{i}$$^{\mathsf{a}}$ & PA$^{\mathsf{b}}$  & $M_{dyn}$$^{\mathsf{c}}$& $\sigma$$^{\mathsf{d}}$ &  Type$^{\mathsf{e}}$ & Companion$^{\mathsf{f}}$ & Class$^{\mathsf{g}}$\\
{} 		& (km~s$^{-1}$) &(degrees)& $10^{9} M_{\odot}$	& (km~s$^{-1}$) 	& {} 			&  \\
\hline	
NGC 7673			& $43$	&168&$0.90\pm0.26$	& $44$	   	& Sa	  		& 	 Yes 	  	& RD\\	   
NGC 7714          		& $98$	&45&$6.03\pm0.75$	& $75$	   	& Sb	  		&        Yes 	   	& RD\\
NGC 6052	 		& $212$	&26&$25.1\pm2.3$		& $98$	   	& Sc	  		&        Yes 	   	& RD\\
NGC 469            		& \ldots 	&\ldots&				& $42$	   	& S0	  		& 	  \ldots	& PR \\
UCM 0000     	 		& $136$	&21&$15.9\pm1.4$		& $192$	   	& Sa	  		& 	  \ldots	& RD \\
UCM 0156     	 		& \ldots  	&\ldots&				& $57$	   	& Sb	  		& 	  \ldots	& PR\\
UCM 1428	 		& \ldots	&\ldots&				& $64$	   	& Irr	  		& 	  \ldots	& CK\\
UCM 1431	 		& $132$	&40&$10.9\pm1.2$		& $95$	   	& Sb	  		& 	  \ldots	& RD\\
UCM 1648	 		& \ldots	&\ldots&				& $47$	   	& Sa	  		& 	  \ldots	& PR \\
UCM 2250     	 		& \ldots 	&\ldots&				& $88$	 	& Sa	  		& 	  \ldots	& CK \\
UCM 2258    	 		& \ldots	&\ldots&				& \ldots  	 	& Sc	  		&        Yes	   	& RD/PR\\
UCM 2317     	 		& $144$	&274&$30.9\pm2.3$		& $86$	 	& Sa	  		&        \ldots	& RD\\
UCM 2327$^{\mathsf{h}}$& \ldots 	&\ldots&				& $47$	  	& Sb \& S0	& 	 Yes 	  	& CK\\
SDSS 1134	 		&  $224$	&70&$29.2\pm2.6$		& $193$	  	& Sa	  		& 	 Yes 	   	& RD\\
SDSS 1507	 		&  $117$	&40&$5.09\pm0.77$	& $38$	  	& Sd	  		&        \ldots	& RD\\
SDSS 1605			& \ldots	&\ldots&				& $30$	   	& Irr	 		& 	 Yes 	   	& CK\\
SDSS 1652			& \ldots   	&\ldots&				& $50$	   	& Sc	  		& 	  \ldots	& PR  \\
SDSS 1703a			& \ldots 	&\ldots&				& $67$		& Irr	  		& 	 Yes   	& CK \\
SDSS 1703b	 		& \ldots    	&\ldots&				&  $31$    		& Sd 		&      	 Yes    	& PR\\
SDSS 1710	 		& $52$	&122&$0.94\pm0.22$	& $58$		& Sd	  		& 	  \ldots	& RD\\
SDSS 2327	 		& $79$	&49&$3.48\pm0.53$	& $46$		& Sc	  		& 	  \ldots  	& RD\\
\hline
Averages$^{\mathsf{i}}$ & $124\pm60$& $85\pm76$&$13\pm11$& $72\pm46$ &\ldots & \ldots \\
\hline
\multicolumn{8}{l}{$^{\mathsf{a}}$ As stated in Section 3, without considering inclination uncertainties,} \\
\multicolumn{8}{l}{~~we estimated an average uncertainty of $\pm7$~km~s$^{-1}$ in $v_{rot}$}\\
\multicolumn{8}{l}{$^{\mathsf{b}}$ Estimated PA; $^{\mathsf{c}}$ $M_{dyn}=\frac{v_{rot}^2R_e}{G}$} \\
\multicolumn{8}{l}{$^{\mathsf{d}}$ As stated in Section 2, for high S/N Gaussian profile fits,}\\ 
\multicolumn{8}{l}{~~the uncertainty is $\pm$2\% to $\pm$10\% in $\sigma$ depending on S/N}\\
\multicolumn{8}{l}{$^{\mathsf{e}}$ Nasa Extragalactic Database: NGC Galaxies, SDSS 1134, and SDSS 1507;}\\
\multicolumn{8}{l}{~~UCM Catalog: UCM Galaxies; the rest were classified by eye}\\
\multicolumn{8}{l}{$^{\mathsf{f}}$ NASA Extragalactic Database; $^{\mathsf{g}}$ Kinematic Classification}\\
\multicolumn{8}{l}{$^{\mathsf{h}}$ As explained in the text UCM 2327n and UCM 2327s are considered as one in our analysis}\\
\multicolumn{8}{l}{$^{\mathsf{i}}$ Mean and standard deviation of the sample}\\
\end{tabular}
\label{table:statistics1}
\end{table*}

\begin{table*}
\centering
\caption{Multiple Kinematic Components}
\vspace{0.5cm}
\begin{tabular}{lccccccc}
\hline
{Name} 	 & Spectral Components$^{\mathsf{a}}$  & {}	& {} & {} 	& Spatial Components$^{\mathsf{b}}$	& {} 	 	\\
{} 	  	& $A_2$ 	& $\overline{{\Delta}I_{max}}$ & \multicolumn{2}{c}{$\overline{{\Delta}\lambda_c}$}	& $N$ 	& $\overline{\Delta{v}}$& $A_1$	\\
{} 		& (\%)	& (\%)			 	        & ({\AA}) &  (km~s$^{-1}$)						& {}		&  (km~s$^{-1}$)				& (\%) 	\\
\hline	
NGC 7673		& \ldots	  	& \ldots 	   	& \ldots&			& 2	 	& $35\pm7$	& $3.0\pm0.7$ \\	   
NGC 7714		& $6.8\pm0.7$	& $45\pm9$	& $2.3\pm0.4$&$136\pm3$	& 1     	& $63\pm14$	& $5.8\pm0.6$\\
NGC 6052		& $20\pm1$	& $43\pm6$ 	& $3.0\pm0.2$&$177\pm2$	& \ldots	& \ldots	   	& \ldots \\
NGC 469            	&  \ldots	  	& \ldots	   	& \ldots&			& \ldots	& \ldots	   	& \ldots \\
UCM 0000     	 	&  $22\pm4$	& $41\pm12$ 	& $5.3\pm1.3$&$311\pm10$	& \ldots	 & \ldots	   & \ldots \\
UCM 0156     	 	& $13\pm6$	& $48\pm35$	& $3.3\pm0.6$&$195\pm5$  	& \ldots	 & \ldots	   & \ldots \\
UCM 1428	 	& $3\pm3$	& $71\pm10$	& $1.9\pm0.1$&$112\pm1$  	& \ldots	 & \ldots	   & \dots \\
UCM 1431	 	& \ldots	  	& \ldots	   	& \ldots&\ldots 	  		& \ldots	& \ldots	   &\ldots\\
UCM 1648	 	& \ldots	  	& \ldots	   	& \ldots&\ldots	  		& \ldots	&\ldots 	   &\ldots\\
UCM 2250     		& \ldots 	  	& \ldots	   	& \ldots&\ldots	  		& \ldots	& \ldots	   &\ldots\\
UCM 2258    	 	& \ldots 	  	& \ldots	   	& \ldots&\ldots	 		& \ldots	& \ldots	   &\ldots\\
UCM 2317     	 	& \ldots 	 	& \ldots	   	& \ldots&\ldots	  		& 2	 & 	$193\pm90$   & $4.0\pm0.6$\\
UCM 2327    	 	& \ldots 	 	& \ldots	   	& \ldots&\ldots	  		& \ldots	& \ldots	   &\ldots \\
SDSS 1134	 	& \ldots 	 	& \ldots	   	& \ldots&\ldots	  		& \ldots	& \ldots	   &\ldots \\
SDSS 1507	 	& \ldots 	 	& \ldots   	   	& \ldots&\ldots	  		& \ldots	& \ldots	   &\ldots\\
SDSS 1605	 	&  \ldots	  	& \ldots	  	& \ldots&\ldots	  		& \ldots	& \ldots	   &\ldots\\
SDSS 1652	 	&  \ldots 	  	&\ldots 	   	& \ldots&\ldots	  		& \ldots	& \ldots	   &\ldots\\
SDSS 1703a	 	& \ldots 	  	& \ldots	   	& \ldots&\ldots	  		& \ldots	& \ldots	   &\ldots\\
SDSS 1703b	 	& \ldots      	& \ldots     		& \ldots&\ldots 	  		& \ldots     	& \ldots	   &\ldots \\
SDSS 1710	 	& $9\pm4$ 	& $73\pm24$	& $1.8\pm0.5$&$106\pm4$	& \ldots	 & \ldots	   & \ldots \\
SDSS 2327	 	& \ldots 	  	& \ldots   		& \ldots&\ldots	 		& \ldots  	&\ldots	   &\ldots \\
\hline
Averages$^{\mathsf{c}}$	&$12\pm7$& $54\pm15$ & $2.9\pm1.3$ &$170\pm76$& $1.7\pm0.5$ & $97\pm84$ & $4.3\pm1.4$ \\
\hline

\multicolumn{8}{l}{$^{\mathsf{a}}$ $A_2$: extension of the spectral components in percentage over the area of the galaxy; $\overline{{\Delta}I_{max}}$: average intensity between components;}\\
\multicolumn{8}{l}{~~$\overline{{\Delta}\lambda_c}$: average distance between components}\\
\multicolumn{8}{l}{$^{\mathsf{b}}$ $N$: number; $\overline{\Delta{v}}$: average velocity with respect to the surroundings;}\\
\multicolumn{8}{l}{~~$A_1$: extension of the spatial components in percentage over the area of the galaxy}\\
\multicolumn{8}{l}{$^{\mathsf{c}}$ Mean and standard deviation of the sample}\\
\end{tabular}
\label{table:statistics2}
\end{table*}

We find  48\% RDs, 28\% PRs, and 24\% CKs (see Table \ref{table:statistics1}). Note that we were unable to classify UCM 2258, for which we do not have velocity width measurements. If we do not consider NGC 7673, NGC 7714, and UCM 2317, galaxies whose rotating nature was found only after carefully investigating our data and removing asymmetries introduced by spatially resolved independent kinematic components, as discussed below, we would find 33\% RDs, 28\% PRs, and 39\% CKs, which is in good agreement with \citet{yang08}. Note that such components would have not been possible to identify at intermediate- and high-redshift. The large variation found when trying to classify the objects in our sample objectively into the different groups defined by \citet{yang08} leads us to be cautious about any interpretation that may follow the classification of such objects at intermediate redshift. 

This shows how the technical limitation of distant observations may play a role in the kinematic classification of these galaxies. This ambiguity may be particularly important when trying to estimate dynamical masses of distant LCBGs using the integrated velocity width. The presence of different asymmetries in the velocity maps might be important contributors to the velocity widths, along with the overall rotation of the system. If we do not consider this, the incorrect dynamical mass estimates could support the wrong evolutionary scenario.  

\subsection{Kinematic Properties}\label{kinprop}

Table~\ref{table:statistics1} lists an estimation of $v_{rot}$ (where possible) and a measurement of the integrated velocity width for each galaxy as a result of summing all the available spectra. We estimated for the latter an average uncertainty of $\pm2\%$, as stated above, for high S/N Gaussian profile fits. Furthermore, the galaxy morphological type, and whether or not a known companion is present, are also listed. When the morphological type was not available in the literature, a classification was performed by eye using the rest of our sample as a reference. We consider the presence of companions when either one of the two following statements apply: 
(i) companions have previously been identified in the literature and are identified as such by the NASA Extragalactic Database\footnote{The NASA/IPAC Extragalactic Database (NED) is operated by the Jet Propulsion Laboratory, California Institute of Technology, under contract with the National Aeronautics and Space Administration.}; and (ii) companions are found in our $74\times65$~arcsec$^2$ FOV (i.e., $28\times25$~kpc$^2$ at the average redshift of our sample) with a redshift within the maximum and minimum velocities of the main target (e.g., SDSS 1703a and SDSS 1703b).

\begin{figure*}
  \begin{center}
\begin{tabular}{ccc}

      \includegraphics[width=5cm]{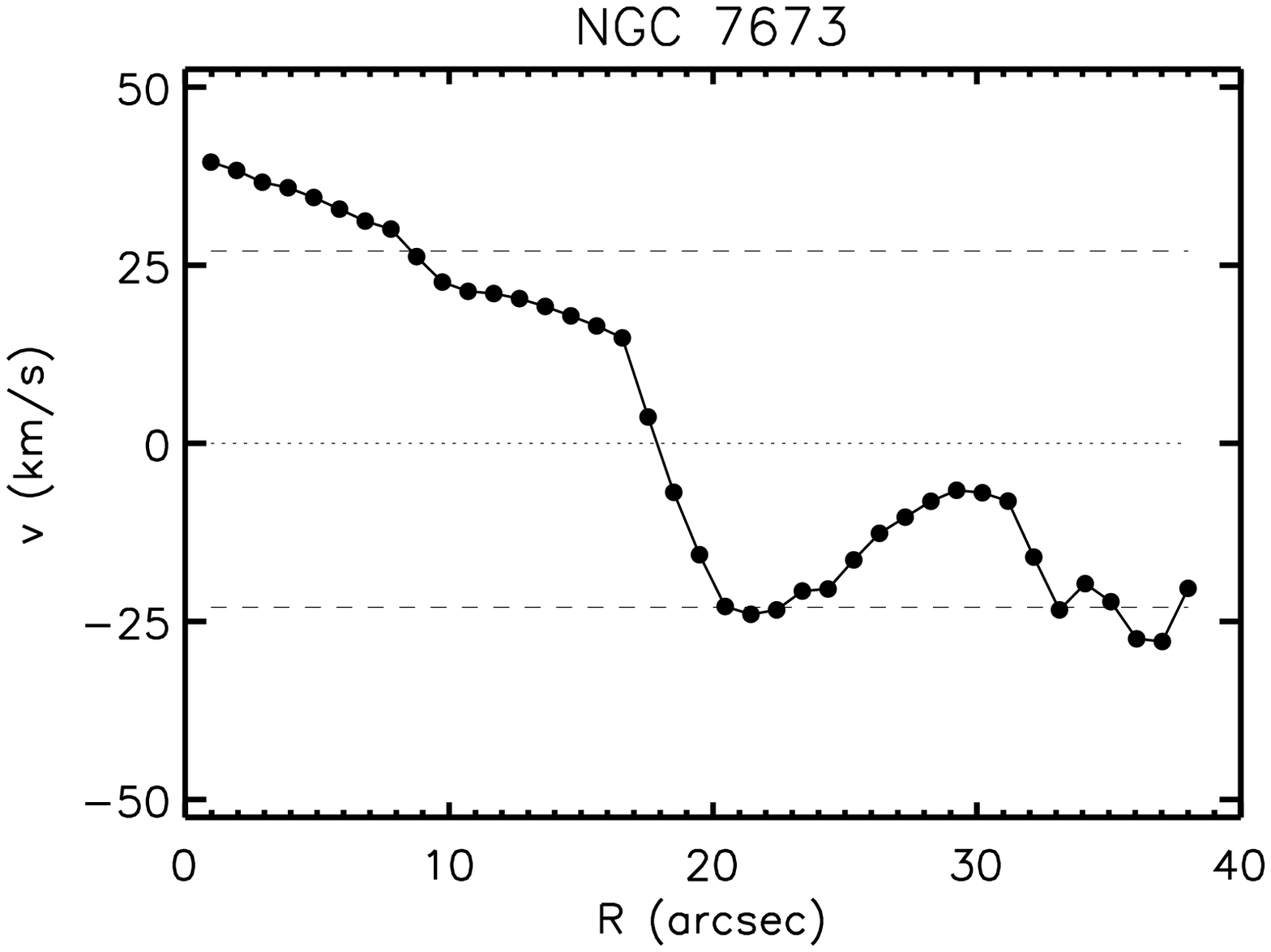} &
      \includegraphics[width=5cm]{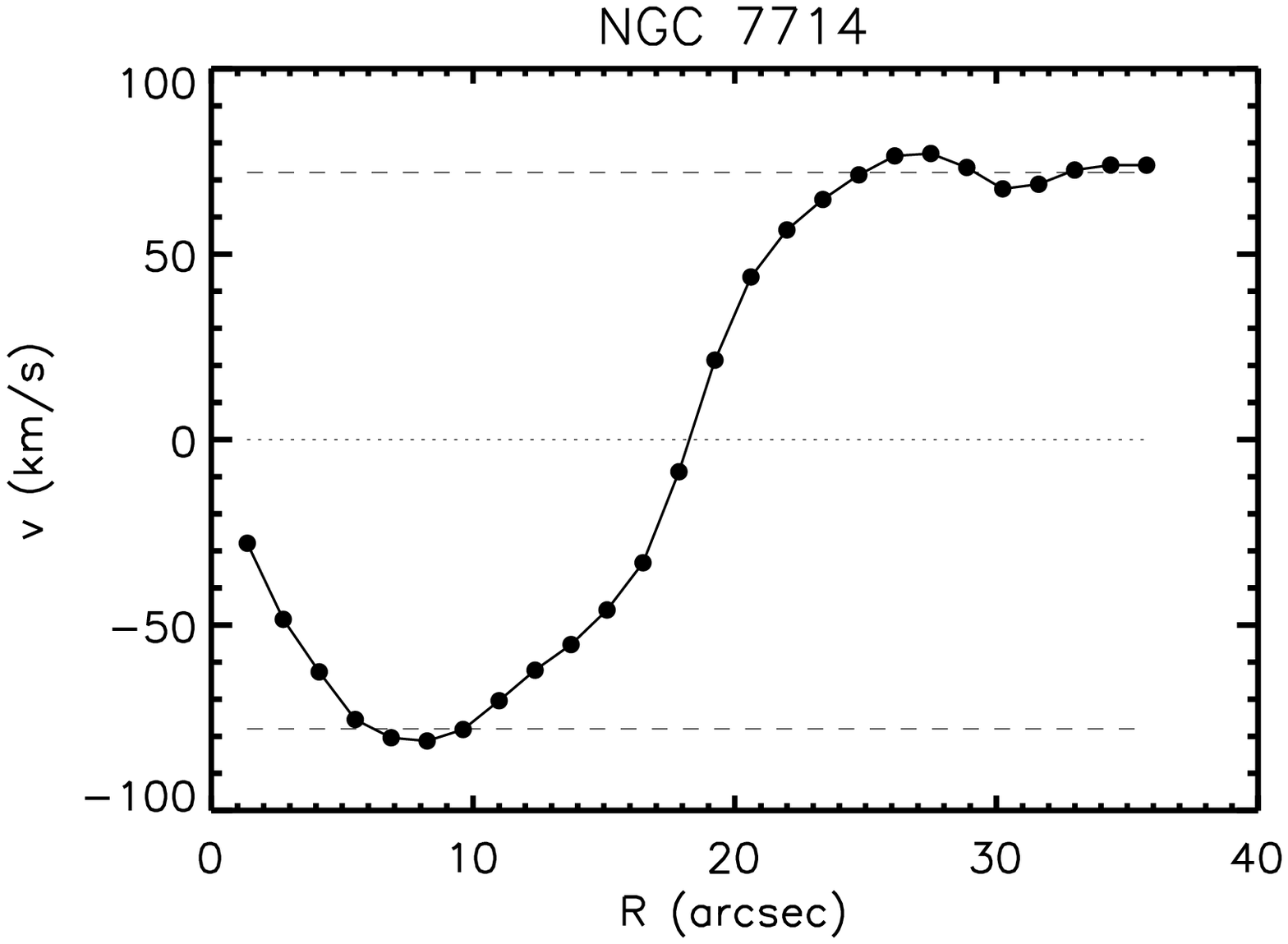} &
      \includegraphics[width=5cm]{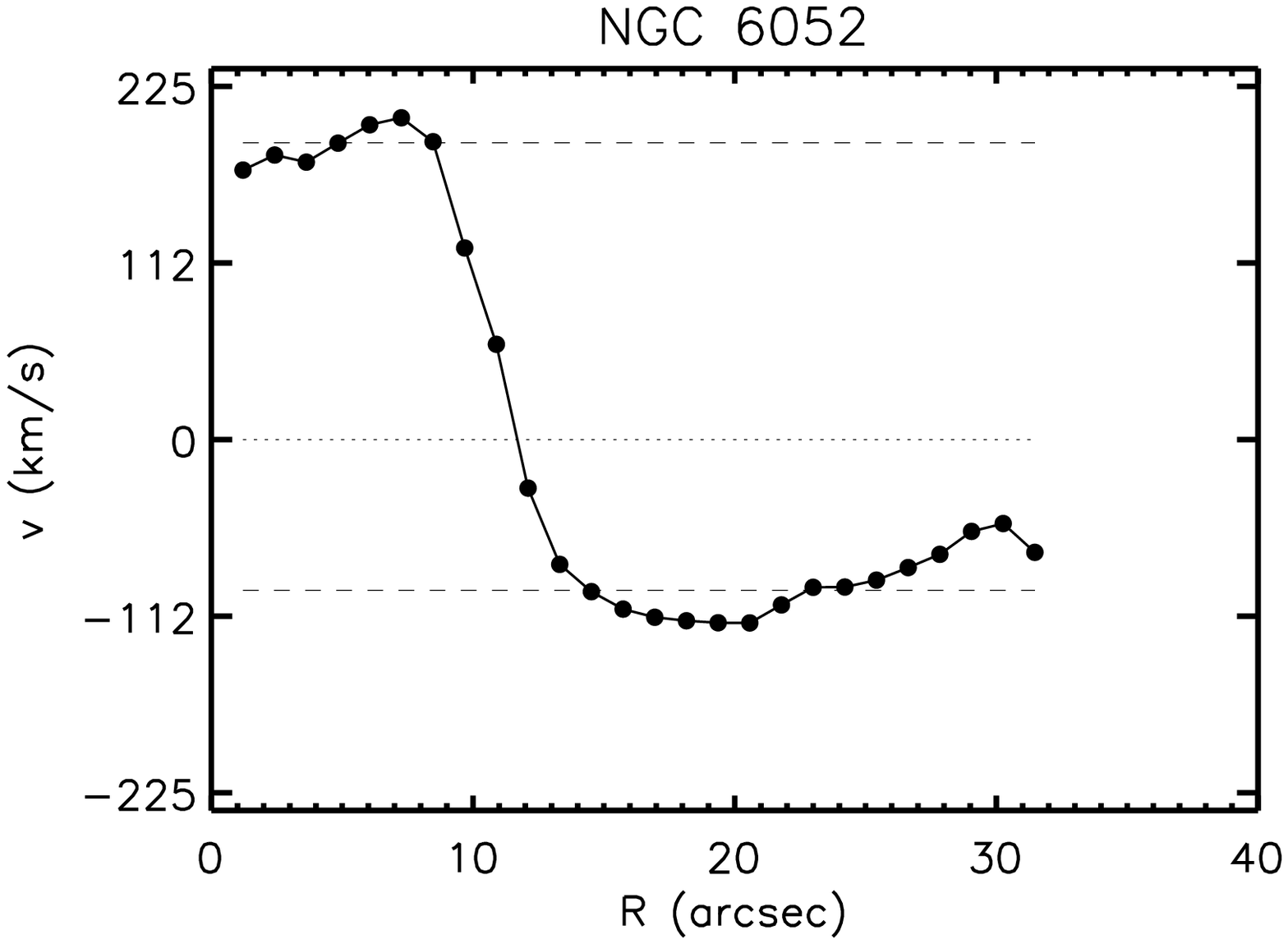} \\
      
      \includegraphics[width=5cm]{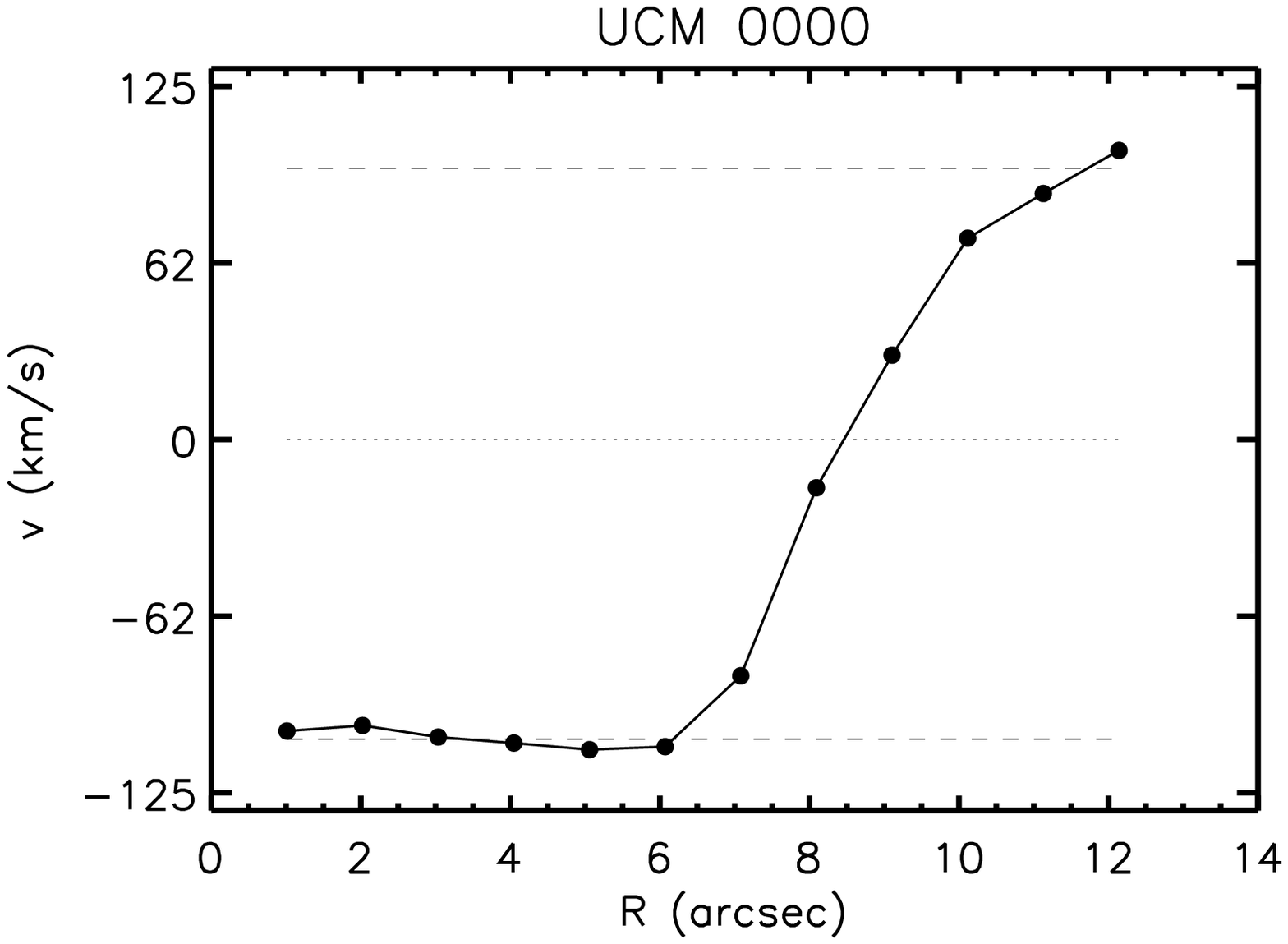} &
      \includegraphics[width=5cm]{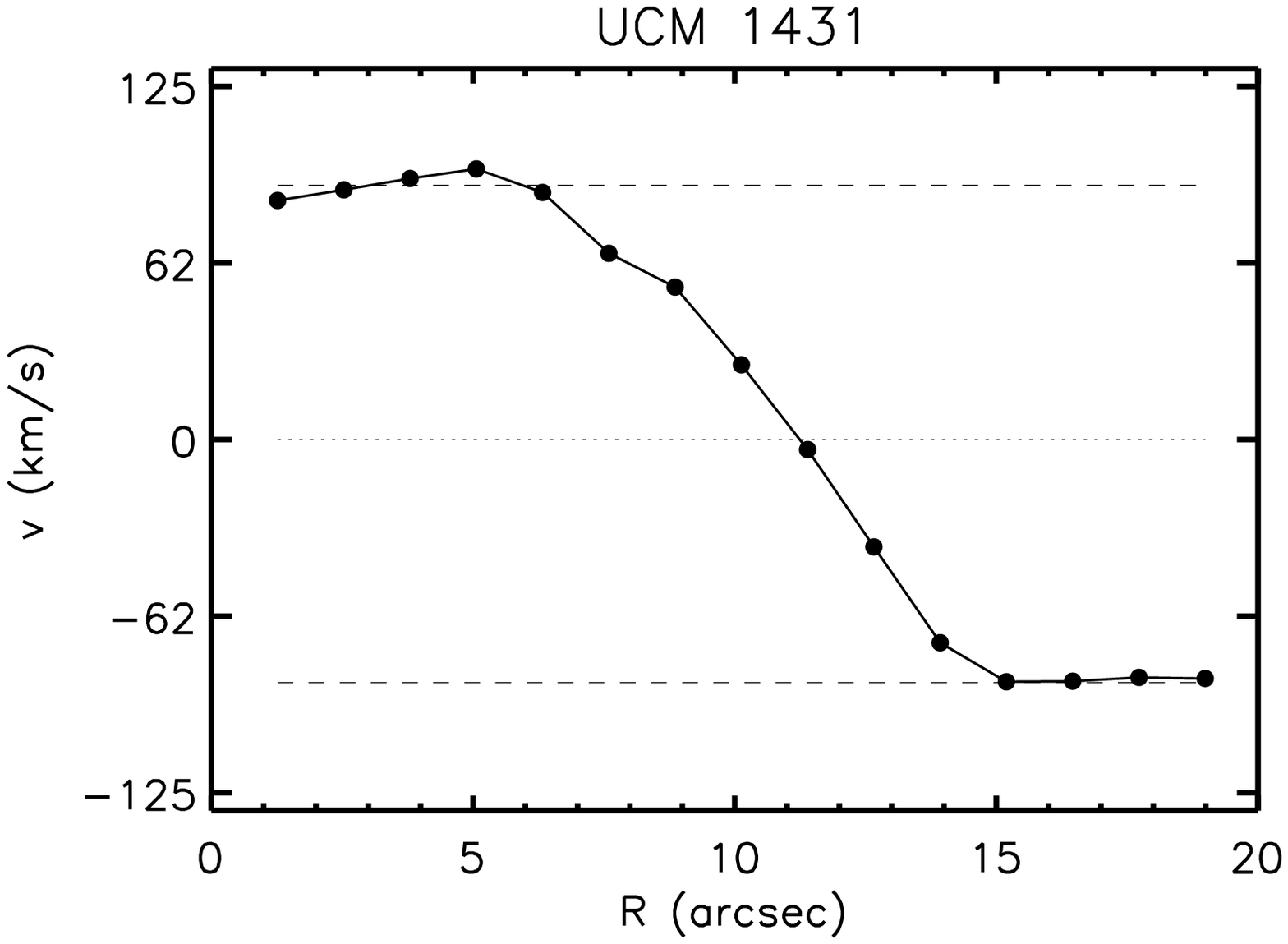} &
      \includegraphics[width=5cm]{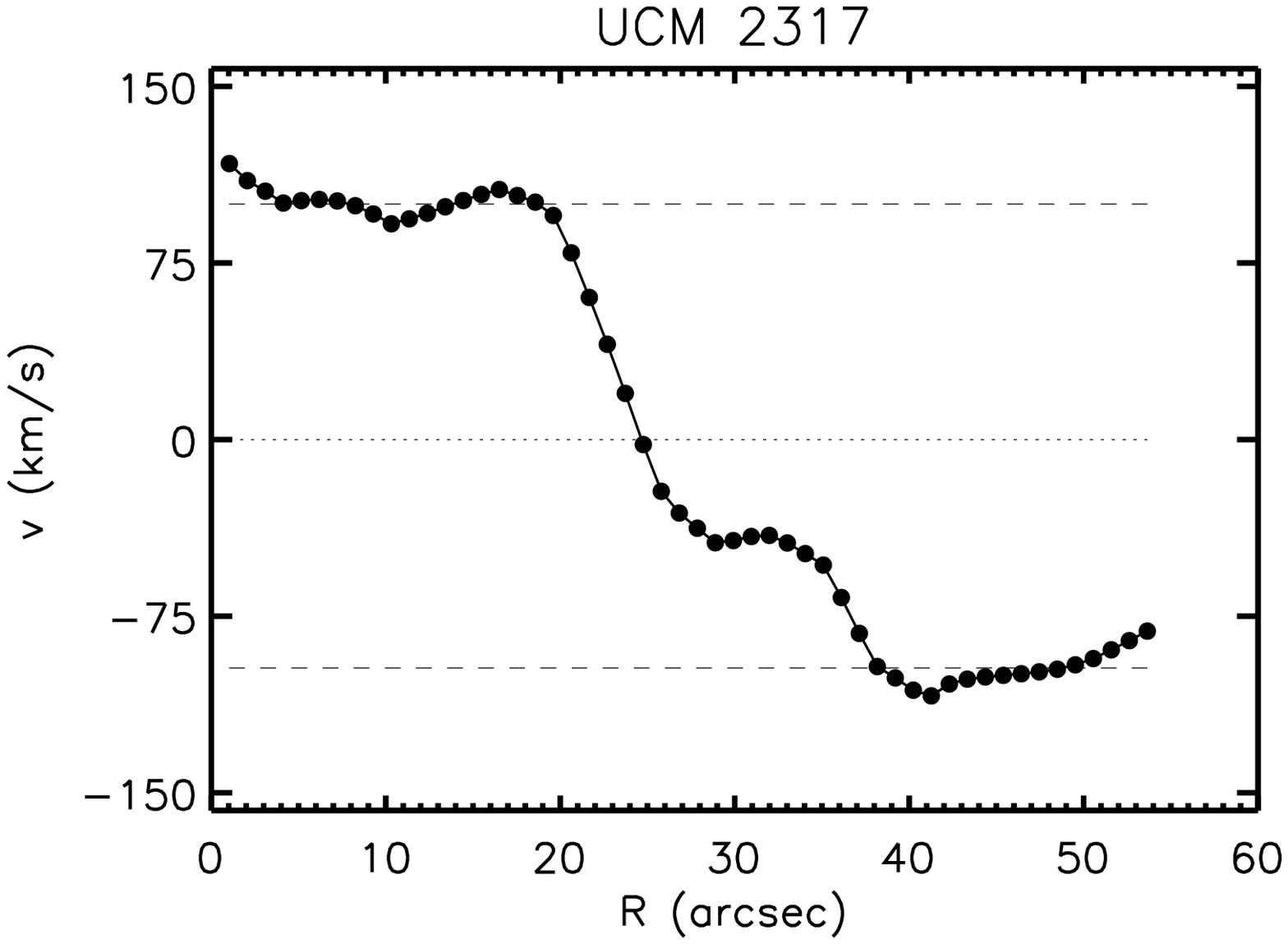} \\

\end{tabular}

\begin{tabular}{cc}

      \includegraphics[width=5cm]{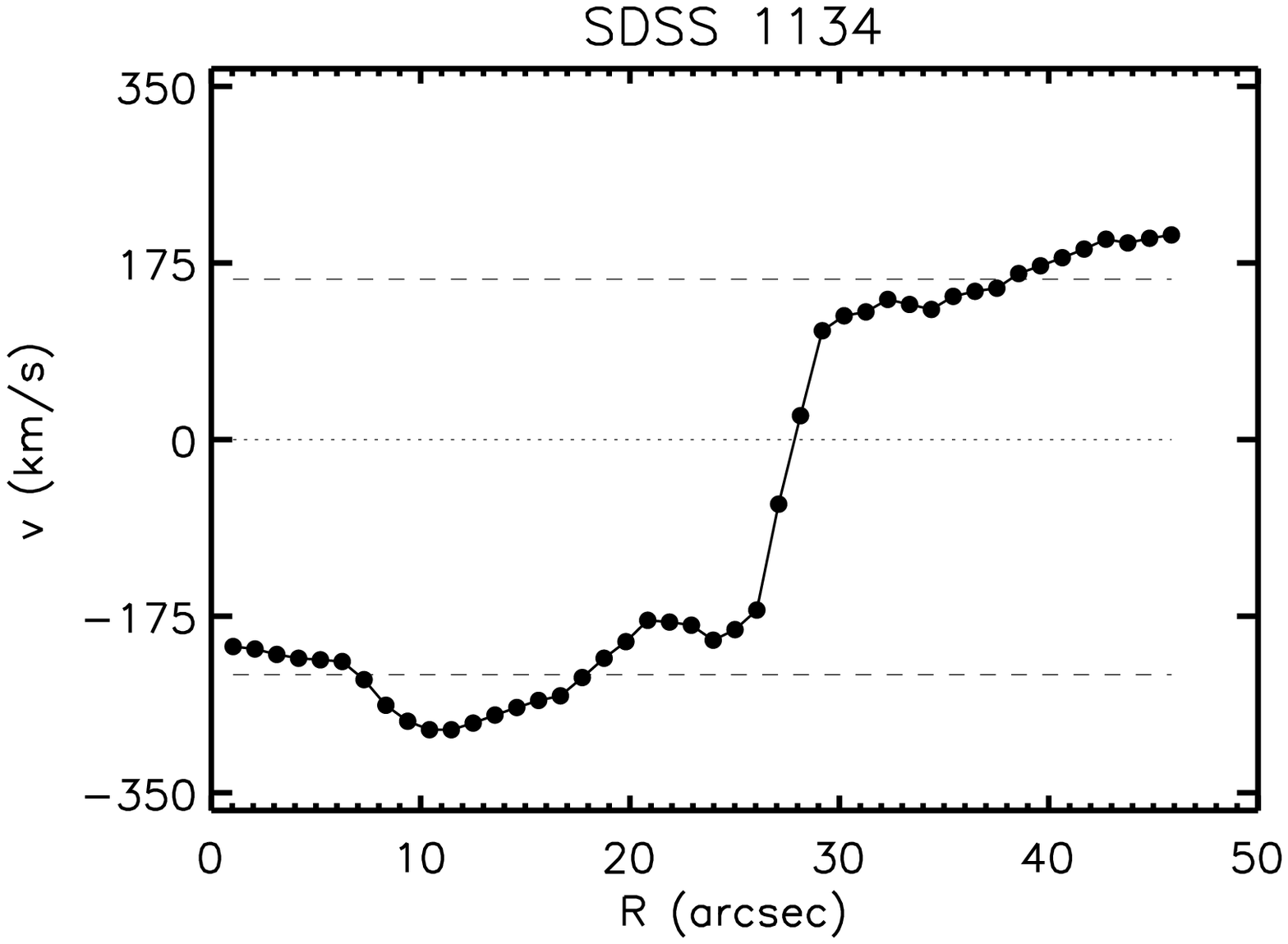} &
      \includegraphics[width=5cm]{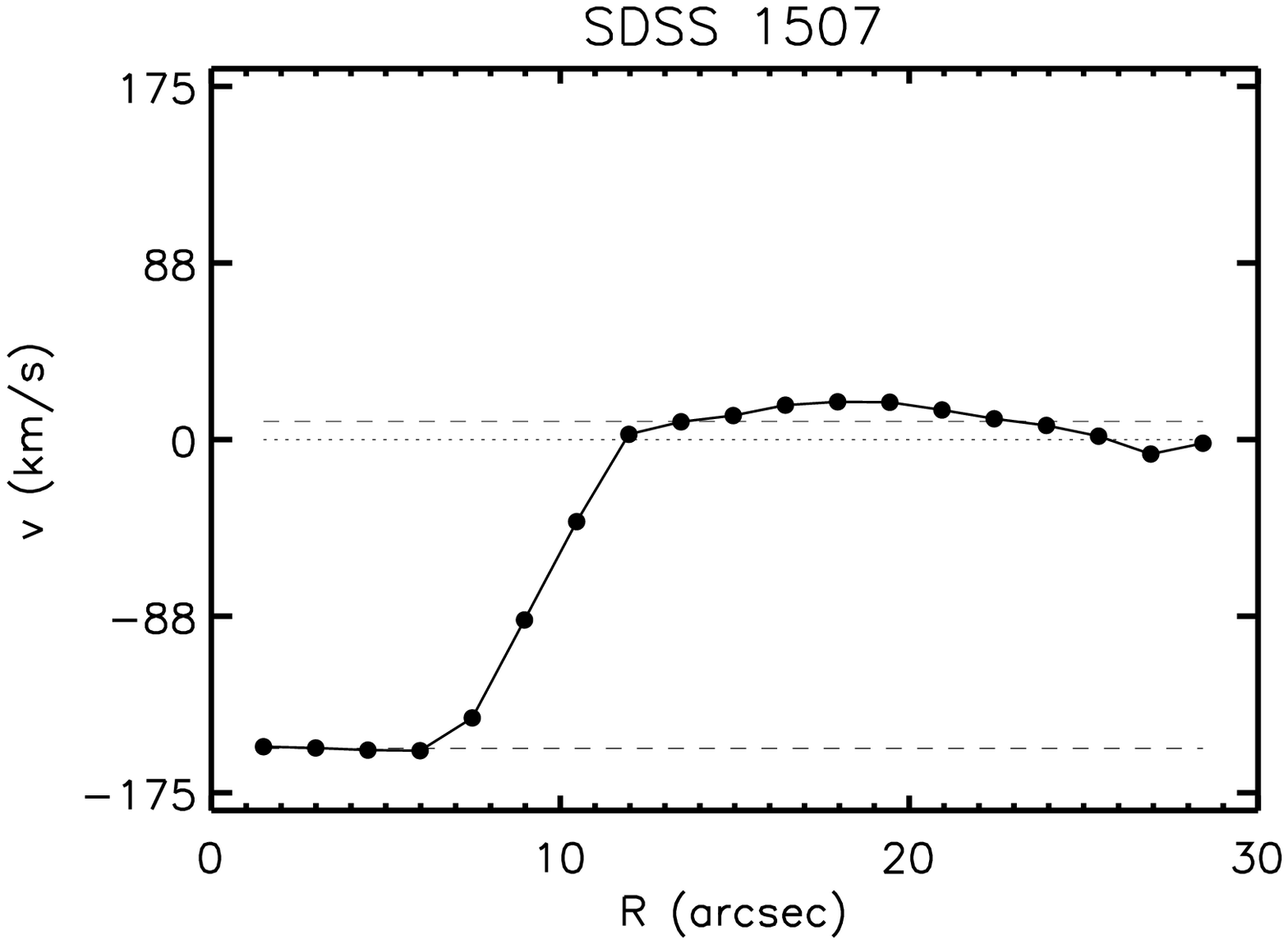} \\

      \includegraphics[width=5cm]{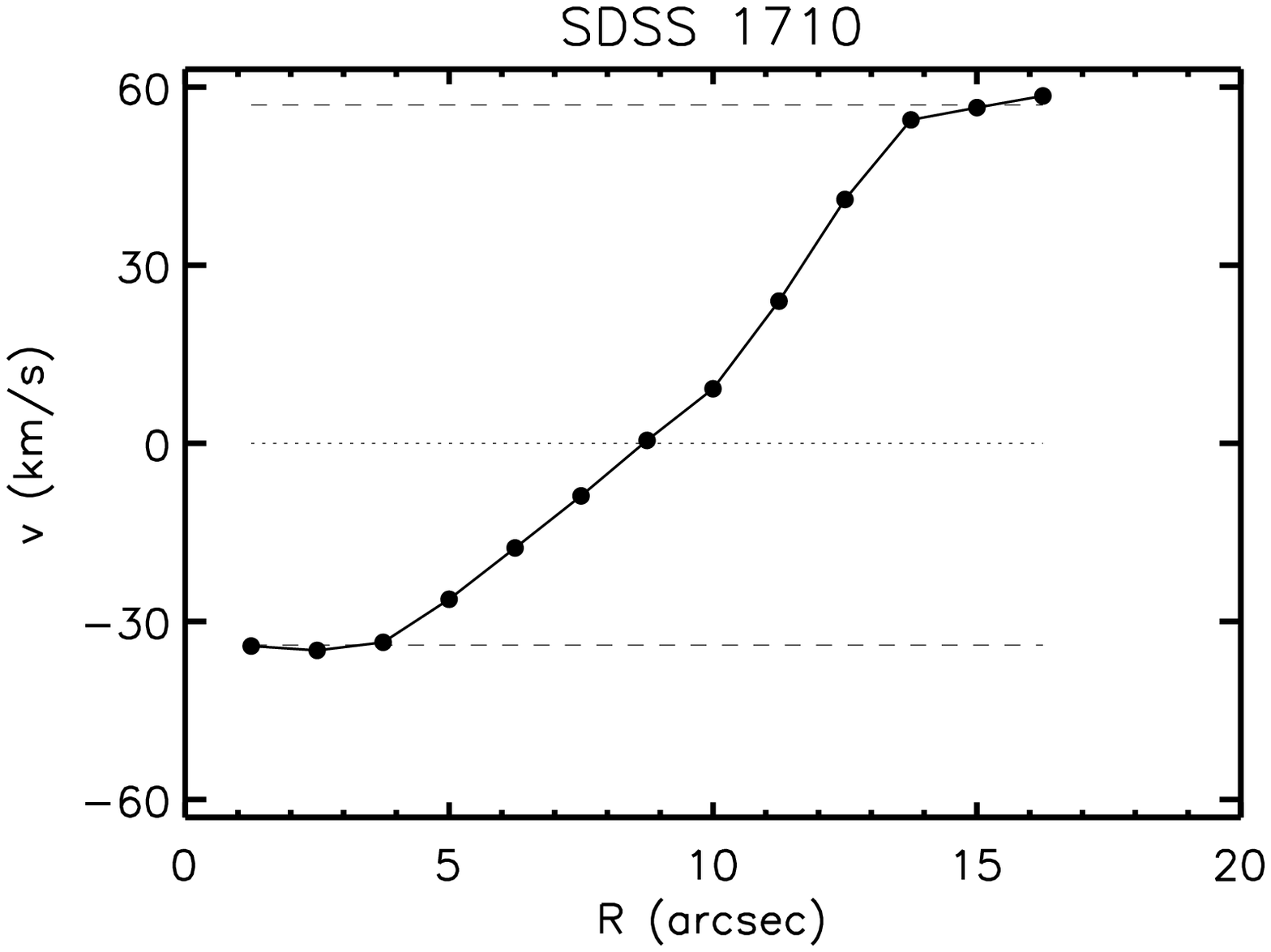} &
      \includegraphics[width=5cm]{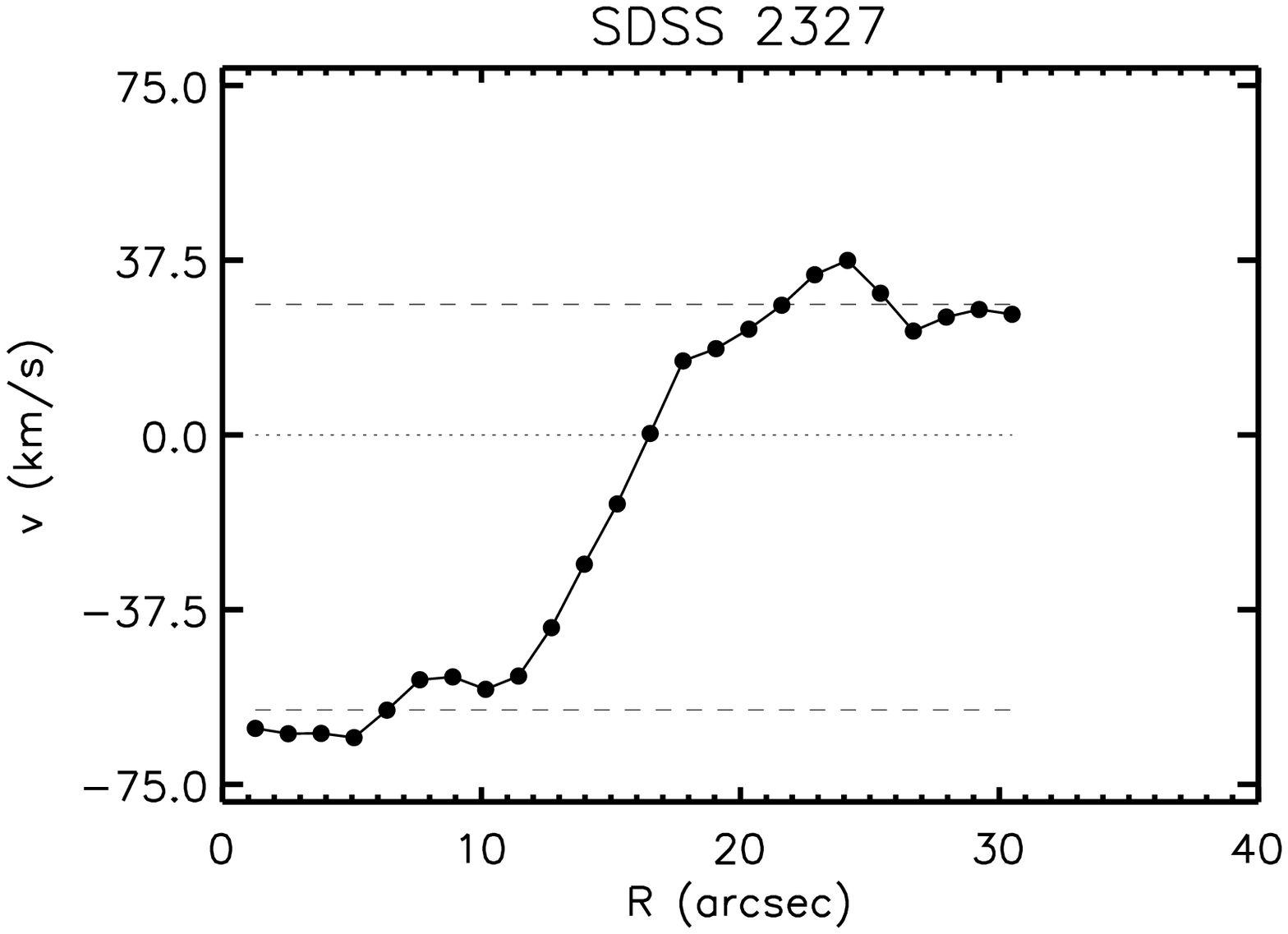} \\

\end{tabular}
         \caption{Rotation curves for UCM 0000, UCM 1431, UCM 2317, SDSS 1134, SDSS 1507, SDSS 2327, SDSS 1710, NGC 7673, NGC 7714, and NGC 6052. For each LCBG, the dashed lines indicate the plateaus, while the dotted lines indicate the velocity of the central velocity width peak, which is taken as a reference. The x-axis indicates the distance in arcsec from one extreme to the other of the major axis of the galaxy. The position of the central velocity width peak corresponds to the intersection of the rotation curve with the dotted line. The spatially resolved kinematic components of UCM 2317 were removed for this analysis. The spatially resolved kinematic components of NGC 7673 (clump on the negative velocity side) and NGC 7714 (arc on the negative velocity side) can be seen easily, although they were not taken into account in the analysis. The typical uncertainties associated with a single data point are 6~km~s$^{-1}$ and 1~arcsec.}
    \label{fig:rotcurv}
  \end{center}
\end{figure*}

An estimation of $v_{rot}$ was performed only for those 10 galaxies classified as RDs. 
For those, galaxy rotation curves were drawn along their major axis as derived from the positions of their kinematic centers, the geometry of their velocity contours, and their position angles (PAs; Table~\ref{table:statistics1}). No fitting was attempted, and $v_{rot}$ was estimated as half the distance between velocity plateaus at both galaxy sides (Figure~\ref{fig:rotcurv}).

Our observational strategy allowed us also to sample the low surface brightness outskirts of these galaxies.
We investigated the effect of this in our determination of $v_{rot}$ by considering $\pm1$ data point in the process of fitting the plateaus and found it is only important for the red plateau of UCM 0000 as can be seen in Figure~\ref{fig:rotcurv}. The average variance of these fits is $\pm1$~km~s$^{-1}$. The uncertainty associated with a single velocity data point is $\pm6$~km~s$^{-1}$. 
A $\pm10^{\circ}$ variation in the position angle produced less than a 1\% change in our
estimate of $v_{rot}$.  From our simple analysis and plateau fits we estimated an average uncertainty of $\pm7$~km~s$^{-1}$ in our measurements of $v_{rot}$ for each galaxy; this does not take into account the effects of the uncertainties associated to the inclination of these galaxies.

To correct $v_{rot}$ for the effects of inclination ($i$), 
we followed the same procedure described in \citep{garland04}.
The SDSS isophotal major and minor axes in the $r$ band (6230 {\AA}) were used. Inclinations from \citet{garland04} were used for NGC 7714 and NGC 6052. The inclinations for NGC 7673 and NGC 469 were taken from \citet{pisano01} and HyperLeda\footnote{http://www-obs.univ-lyon1.fr/hypercat} respectively. Inclinations for the UCM galaxies were obtained from \citet{gonzalez03b}. If an uncertainty of $\pm10\%$ in the inclination is assumed, $v_{rot}$ would range, on average, from $\pm10\%$ in $v_{rot}$.

The dynamical masses of the galaxies, $M_{dyn}=\frac{v_{rot}^2R_e}{G}$, were estimated for the rotating disks in our sample from the value found for $v_{rot}$ and the effective radii listed in Table~\ref{table:properties}. The masses of these objects vary from $\sim$$1\times10^{9}$ to $\sim$$3\times10^{10} M_{\odot}$.

\begin{figure}
\begin{center}
\includegraphics[width=8.cm]{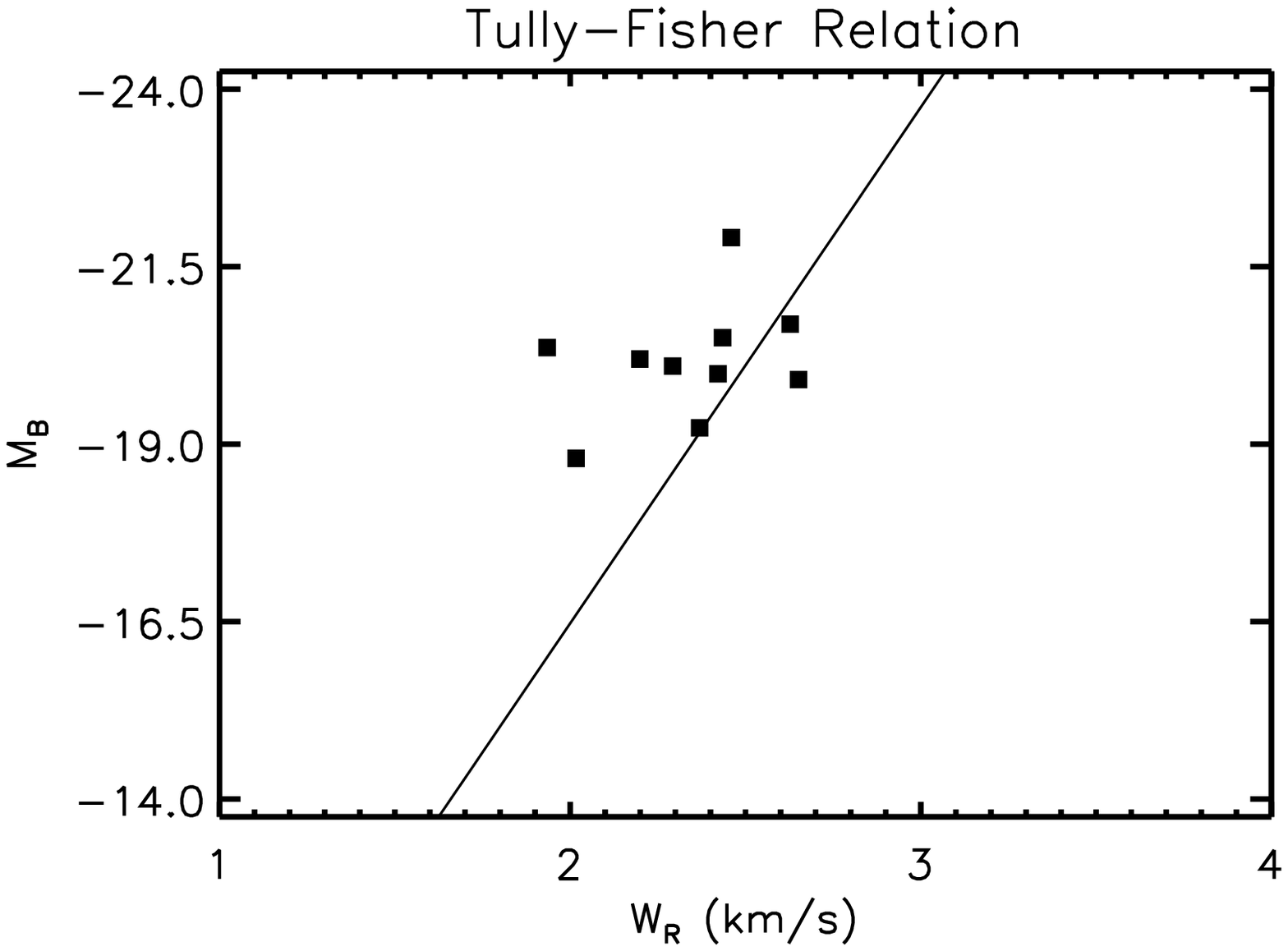}
\caption[Tully-Fisher Relation]{Tully-Fisher relation for a subsample of 10 rotators among our sample of LCBGs (filled squares).The solid line corresponds to a calibration by \citet{tully00}. Symbol size roughly corresponds to 1-$\sigma$ uncertainty on both axes.
\label{fig:tf}}
\end{center}
\end{figure}

Figure~\ref{fig:tf} shows the Tully-Fisher relation for these galaxies compared to a calibration by \citet{tully00} for spirals. Our sample of local LCBGs tends to show larger $M_B$ than the sample of spiral galaxies used for their calibration for a particular $W_R$. Overall, for a particular rotational velocity, they are brighter by half a magnitude. By combining the dynamical masses shown in Table~\ref{table:statistics1} and the absolute magnitudes shown in Table~\ref{table:properties}, we find an average value for $M/L_{B}$ equal to 0.6. Such a lower value is in agreement with those found for late-type galaxies \citep{dickel78}.

A second estimate of $v_{rot}$ can be also made from the measurement of the velocity width by means of 
$W_{20}$ (full width at 20\% of the peak intensity; $W_{20}=3.58\sigma$; $v_{rot}=0.5W_{20}$), or, better, by means of $W_R$ \citep[line width after correcting for turbulent motions;][]{tully85}.
This correction for random motions decreases the line width of the local LCBG by $26-38$~km~s$^{-1}$, depending on the rotational velocity \citep{garland07}. For spiral galaxies $W_R$ is equal to twice $v_{rot}$ 
with a scatter of 9\% \citep{tully85}.
For our sub-sample of rotating LCBGs we find that while this correlation also applies, this is because its dispersion is rather large (i.e. $\sigma=0.55$). Such a dispersion may account for a factor up to $\sim$2 when estimating $v_{rot}$ from velocity widths (with respect to the actual measurement), which translates into a factor of up to $\sim$4 when estimating dynamical masses. This implies that the velocity widths of LCBGs, rather than accounting for the overall rotation of these galaxies, may also significantly account for other kinematic components and, therefore, may not be a good estimate of dynamical mass.

We find that almost half of our sample rotates, with the average ratio $v_{rot}/\sigma\sim2$. It is important to note that in order for the rotating nature of objects such as NGC 7673, NGC 7714, and UCM 2327 to be found, it is necessary first to identify their spatially resolved kinematic components, subtract them, and only then measure $v_{rot}$. 
Local LCBGs typically show integrated velocity widths ranging from $\sim$30\---100~km~s$^{-1}$, in agreement with the range found at intermediate redshift \citep[e.g.,][]{guzman97}. Nevertheless, two objects, UCM 0000 and SDSS 1134, show integrated velocity widths as high as $\sim$200~km~s$^{-1}$. While UCM 0000 owes its high velocity width to the presence of an underlying broad component that is likely caused by an AGN, SDSS 1134's high velocity width is due to its rotating nature, which translates into the only double-horned velocity profile we identify among the integrated spectra of our sample.

Ten of the galaxies in our sample are found to have companions, as discussed in Section 3.  
Half of the galaxies with companions appear to be rotating. It is important to realize that some systems, such as NGC 6052 (a merger of two galaxies), may owe their apparent rotating nature to a projection effect \citep{garland07}. Using IFS, \citet{garcia-lorenzo08} find a similar kinematic behavior, albeit for a smaller FOV, and reach a similar conclusion.

Table~\ref{table:statistics2} lists the presence and properties of spectrally resolved components that might be linked to SN-driven galactic winds \citep[e.g.,][]{marlowe95}; and spatially resolved components that might be linked to infalling galaxies, as discussed by \citet{perez09}. 

Spatially resolved components were identified by studying the velocity map of each galaxy, while independent spectral components were identified by studying the non-Gaussianity of the spectra of each galaxy. The number ($N$), extension ($A_1$; area of the spatially resolved component as compared to the area of the galaxy), and average velocity with respect to their surroundings ($\overline{\Delta{v}}$) are listed for the spatially resolved components. The extension ($A_2$; area of the independent spectral component in comparison with the area of the galaxy), the average intensity between components ($\overline{{\Delta}I_{max}}$), and the average distance between components ($\overline{{\Delta}\lambda_c}$) are listed for the independent spectral components.

Three of the galaxies in our sample (NGC 7673, NGC 7714, and UCM 2317) show spatially resolved kinematic components decoupled from the rest of the galaxy. 
The detection of these components is limited by our ability to find compact kinematic structures with velocities three times the standard error of the velocity larger or smaller than their surroundings, and was possible in RDs since they were identified as kinematically decoupled regions within a rotating background. These kind of detections would have also been possible, for example, within a smooth velocity map in a face-on galaxy. The area of these components was calculated by comparing the number of fibers per dithering in which they were detected with the number of fibers per dithering in which the galaxy was detected. The minimum size we can measure corresponds to one fiber: 2.7~arcsec across or $\sim$1~kpc at the average redshift of our sample.

NGC 7673 shows two spatially resolved kinematic components moving with an average speed of $35\pm4$ km~s$^{-1}$ \citep{perez09}. NGC 7714 shows an arc-like structure that is moving at an average speed of $63\pm5$ km~s$^{-1}$. UCM 2317 shows two extra components located at its core and moving at an average speed of $193\pm34$ km~s$^{-1}$. All five components in these three galaxies are moving away from the observer and falling towards the galaxy, if we assume the galaxy is opaque. This opacity may also explain why we do not see components moving towards the observer and falling towards the galaxy, since they would be behind the galaxy. 

Six of the galaxies in our sample are found to show spectrally resolved kinematic components: double peaks in their emission lines with a typical separation of 2.9 {\AA}). At the average redshift of the galaxies in our sample this implies an offset between the two peaks of about 170~km~s$^{-1}$. Typically, one of the components is twice as intense as the other on average. The detection of these components is limited by the spectral resolution and the S/N of our data. After investigating the residuals of our Gaussian profile fit procedure we found we were able to identify broadening effects down to $\sim$60~km~s$^{-1}$, and offsets between components of $\sim$1 {\AA} ($\sim$60~km~s$^{-1}$ at 5040 {\AA}). The offsets we measure for the different components identified in several galaxies of our sample are twice as large as this lower limit.

The average offset we measure is in good agreement with those measured by \citet{marlowe95} in a local sample of star-forming dwarf galaxies using Echelle spectra and Fabry-Per\'ot images. They interpret these offsets as consistent with SN-driven galactic winds.
These filaments and/or ``superbubbles" have the potential to cause starburst-driven mass loss. As a result, these galaxies cannot retain newly synthesized metals and have low metallicities. For a correlation between these we refer the reader to Castillo-Morales et al. (in preparation). 

\begin{figure}
  \begin{center}
    \centering

      \includegraphics[width=8.cm]{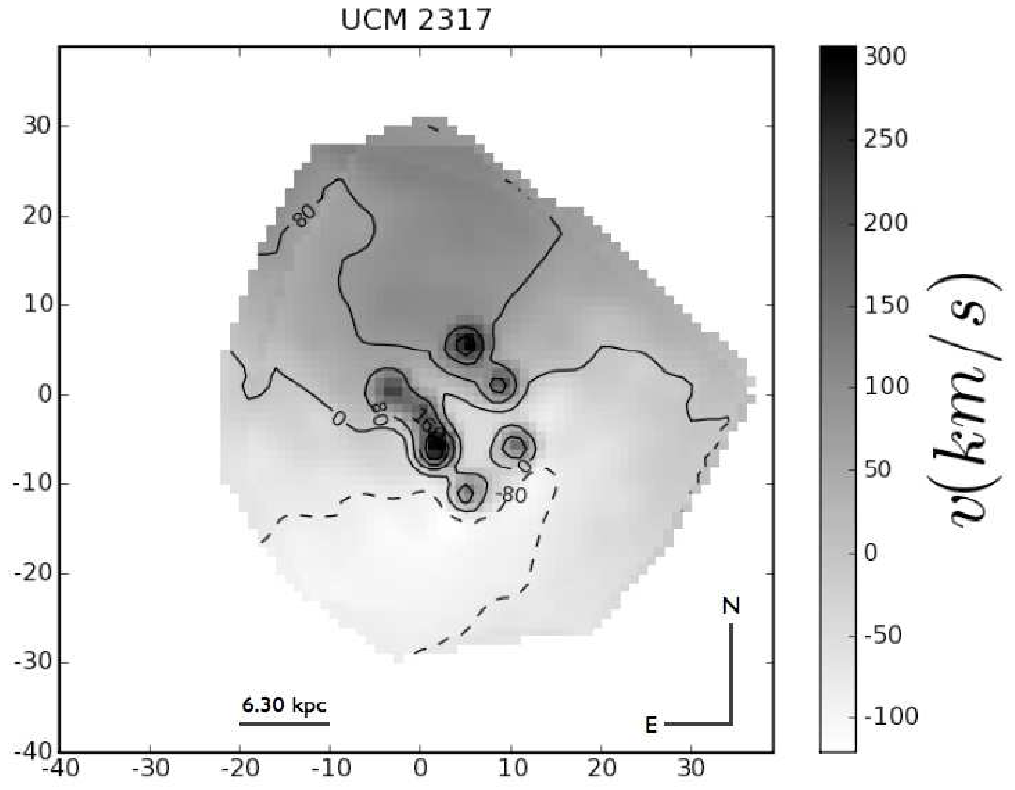} 
      \includegraphics[width=8.cm]{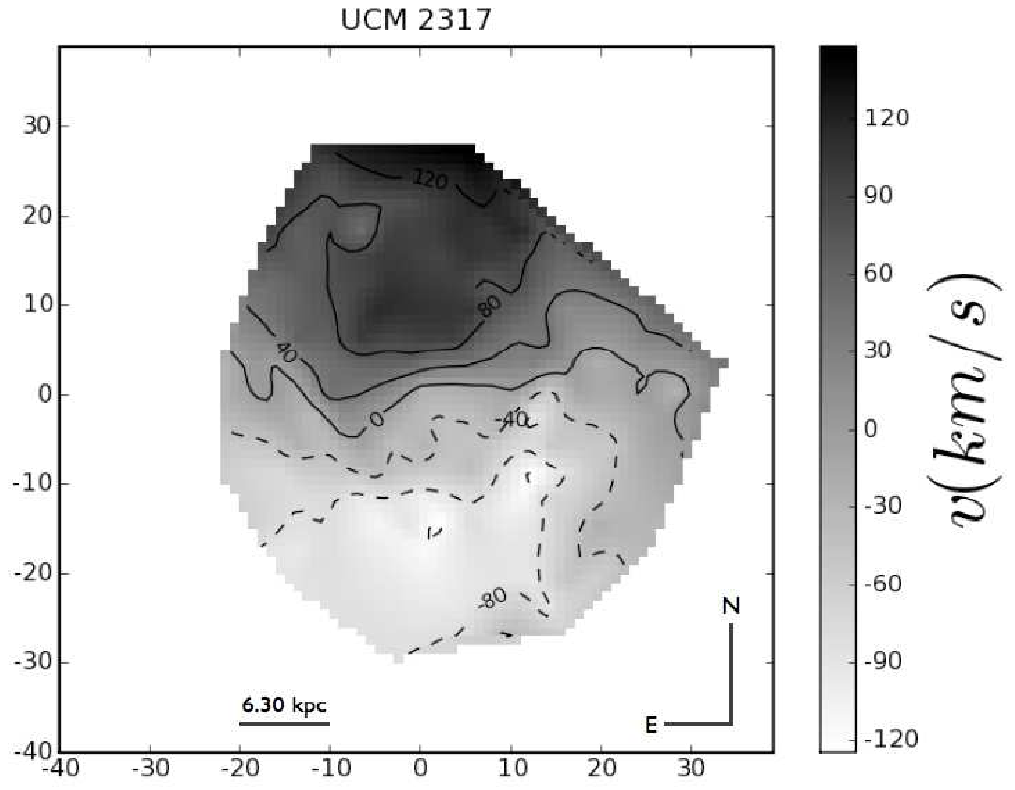}

    \caption[UCM 2317]{Two velocity field maps are shown for UCM 2317 one with ($top$) and one without ($bottom$) the spatial kinematic components, as discussed in the text.}
    \label{fig:ucm2317}
  \end{center}
\end{figure}

Only NGC 7714 shows both spectral and spatial kinematic components. For NGC 7673, NGC 7714, NGC 6052, UCM 0000, UCM 2317, and SDSS 1710 an estimation of $v_{rot}$ was possible while also having either a spectral or spatially resolved kinematic component. The rotating nature of UCM 2317 was actually found after removing these components (Figure~\ref{fig:ucm2317}).

\begin{figure}
  \begin{center}
    \centering

      \includegraphics[width=8.cm]{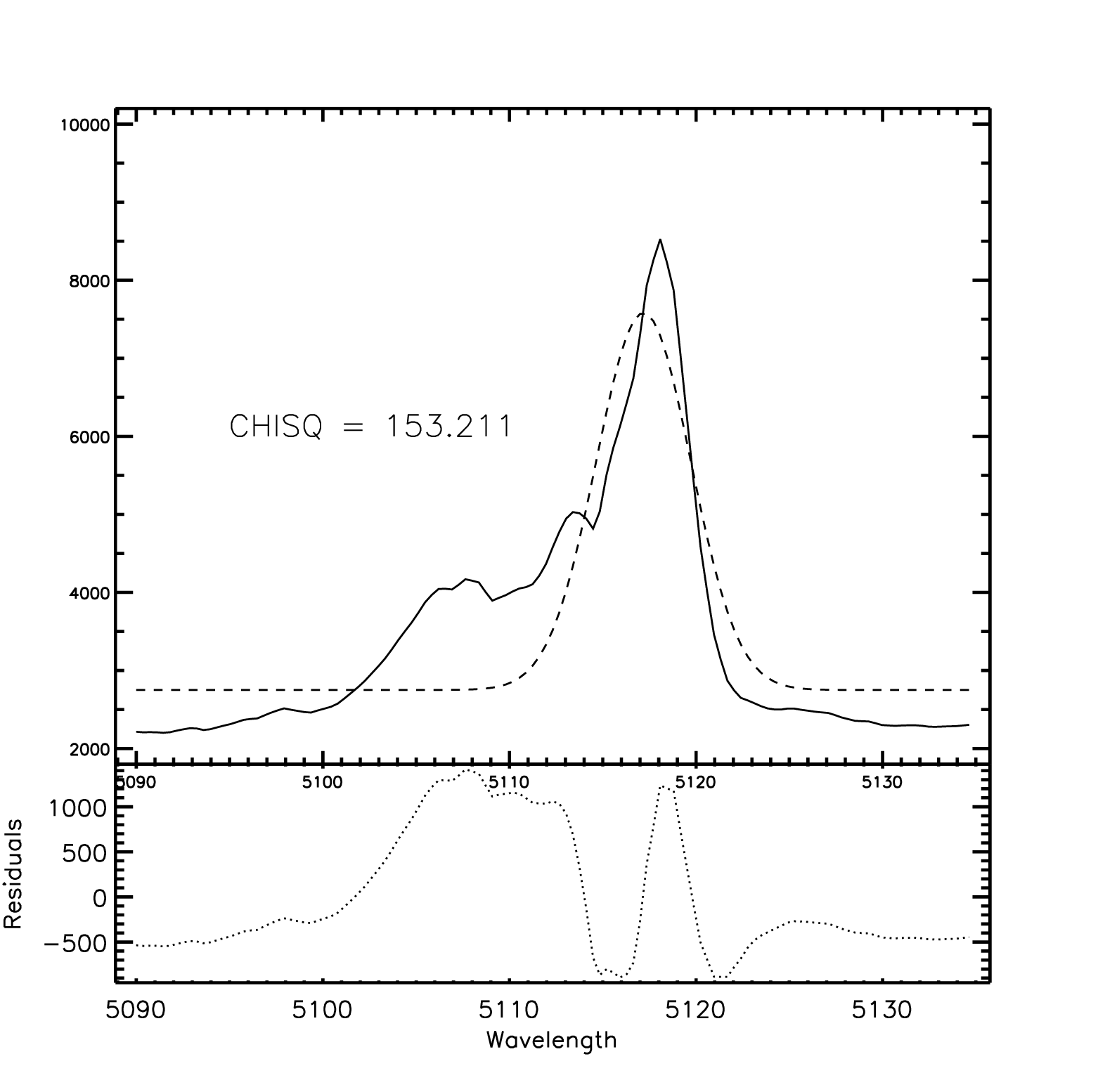} 
      \includegraphics[width=8.cm]{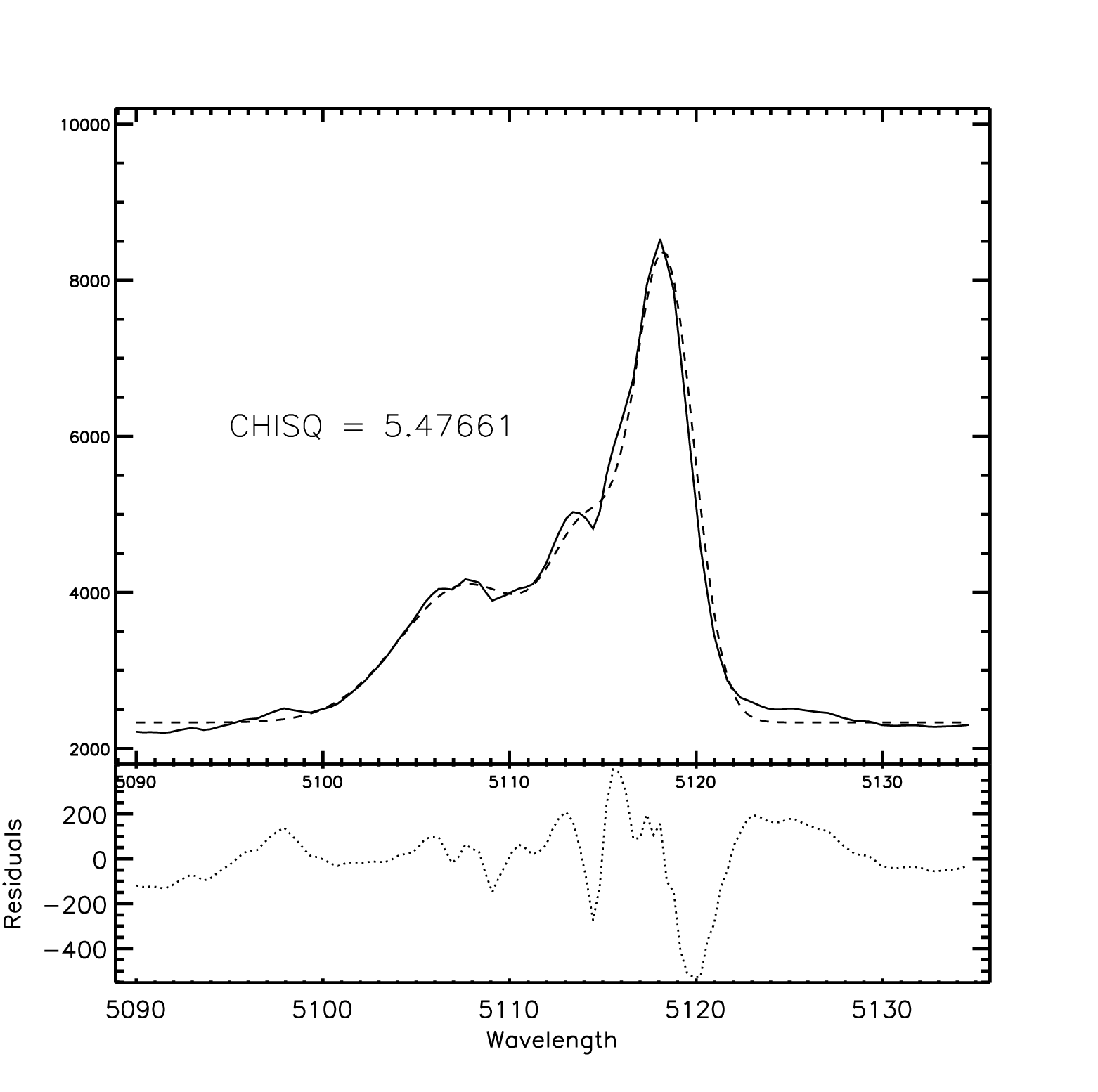}

    \caption[Three Spectral Components on UCM 0000]{Single and triple Gaussian profile fits are shown (dashed line) for this particular spectrum of UCM 0000. The $\chi^2$ (CHISQ) of each fit is shown, as well as the residuals for all the fits in the bottom box of each plot. The flux scale (y-axis) is arbitrary and the wavelength (x-axis) is in {\AA}.}
    \label{fig:169ucm0000}
  \end{center}
\end{figure}

\begin{figure}
  \begin{center}
    \centering
\includegraphics[width=8.cm]{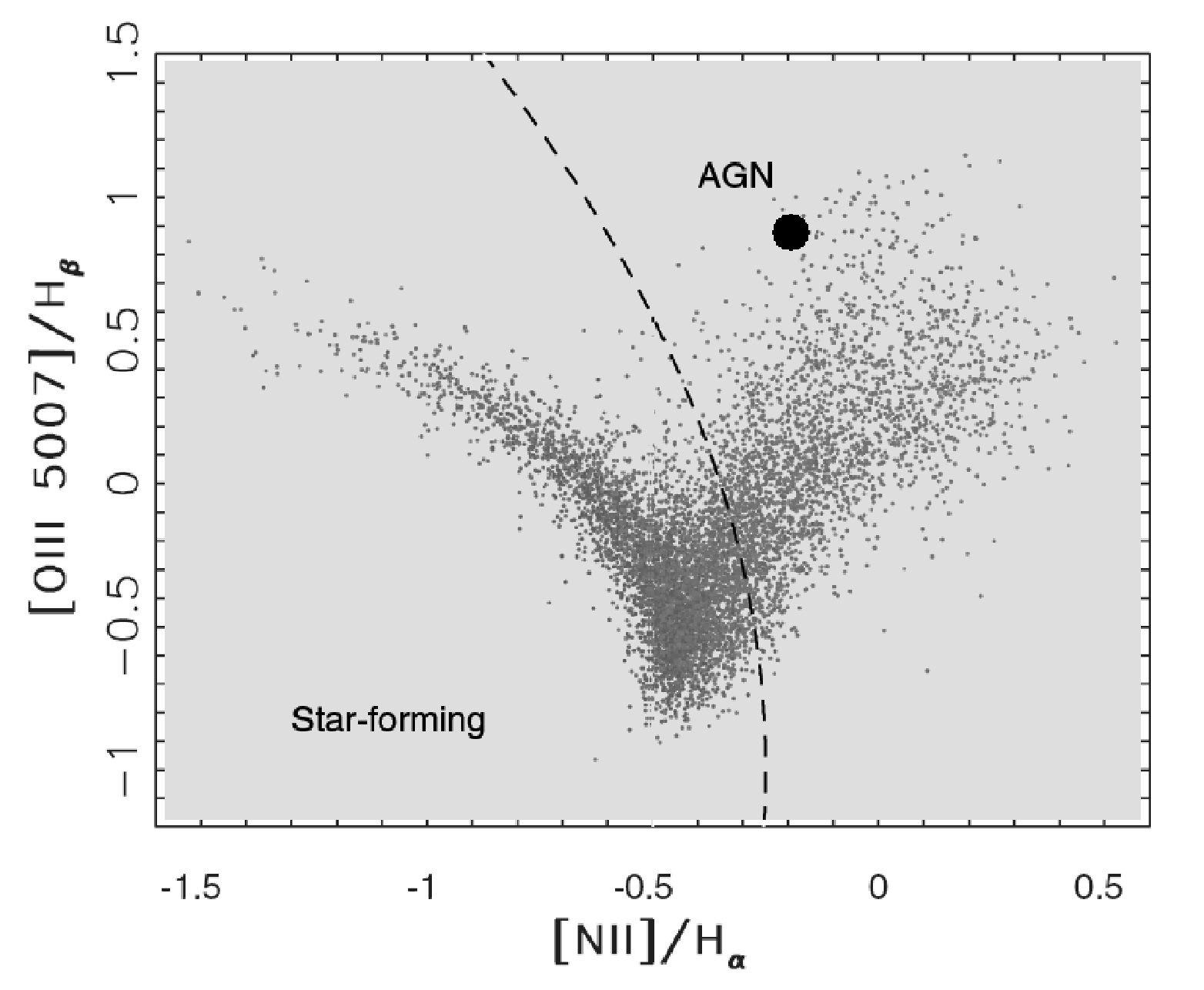}
\caption[Active Galactic Nuclei Diagnostic Diagram]{BPT diagram for SDSS emission-line galaxies (note that the line flux ratios are expressed on a logarithmic scale). UCM 0000 AGN candidate is shown as a large black dot. Emission-line galaxies can be separated into two groups according to their position in the BPT diagram: AGNs, and star-forming, using the separation boundaries outlined by the dashed line. The BPT diagram is taken from \citet{obric06}.}
    \label{fig:agn}
  \end{center}
\end{figure}

UCM 0000 (also MRK 334) shows, not double, but triple spectral kinematic components in 4\% of its area, and a broad component (8.0~{\AA} FWHM, which translates into $\sigma\sim$200~km~s$^{-1}$ at the redshift of the galaxy) in 12\% of its area (Figure~\ref{fig:169ucm0000}; notice that this galaxy shows some sort of extra spectral components in 22\% of its area). This is the only galaxy in our sample showing an obvious broad emission line component. An attempt to investigate the nature of this component was made by calculating the [OIII]$\lambda$5007/H$\beta$, [NII]$\lambda$6584/H$\alpha$, and [SII]$\lambda\lambda$6717,6731/H$\alpha$ ratios for the particular region of the galaxy (i.e., $0.95$, $-0.18$, and $-0.55$~dex, respectively), to study the possible presence of AGN activity. UCM 0000 has been previously classified as a Sy1.8 galaxy by \citet{veron06} and recently studied by \citet{smirnova10}. Figure~\ref{fig:agn} shows the diagnostic line-intensity ratio diagram \citep[also called BPT diagram;][]{baldwin81} for emission-line galaxies in the SDSS \citep[Figure 2 of][]{obric06}. Emission-line galaxies can be separated into two groups according to their position in the BPT diagram: AGNs, and star-forming, using the separation boundaries outlined by the dashed line. These ratios for UCM 0000 are in good agreement with those of an AGN \citep{osterbrock89}. Furthermore, the pressure derived from the [SII]$\lambda\lambda$6717,6731/H$\alpha$ ($\sim0.4$) and [NII]$\lambda\lambda$6548,6584/H$\alpha$ ($\sim0.2$) ratios are in agreement with that of a Seyfert galaxy \citep{rickes08}. Therefore, UCM 0000, is the only galaxy among our sample of 22 LCBGs to show clear evidence for AGN activity. 

Lastly, two out of the three galaxies with spatially resolved kinematic components are found to have a companion, while two out of the six galaxies with spectrally resolved kinematic components are found to have a companion. Furthermore, only one of the galaxies found to have a companion, SDSS 1134, shows an unambiguous rotating nature (i.e., smooth velocity gradient, central velocity width peak, and no spatially resolved kinematic components). If we exclude also NGC 7714, whose spatially resolved kinematic component is a spiral arm, and ignore UCM 2258, for which V1200 configuration data were not available, we are left with five out of five galaxies with both a distorted kinematic behavior and a companion. Out of the 12 galaxies without a companion, six also show a distorted kinematic behavior. In fact, all the galaxies with a companion are either classified as PRs or CKs, or show spatially resolved kinematic components, which denotes the effects companions have on the kinematics of their pairs.

\section{Summary}\label{conclusions}

Velocity and velocity width maps allow us to classify the kinematics of a sample of 22 local LCBGs into three different groups: RD, PR, and CK. We find 48\% RDs, 28\% PRs, and 24\% CKs. 
PRs and CKs have been linked to the presence of both minor and major mergers, or close interactions between galaxies. 
According to this, one half of the galaxies in our representative sample (PRs plus CKs) show kinematics in agreement with either an interaction or a merging scenario. Furthermore, at least two of the galaxies identified as RDs show evidence of being a minor merger. 
At least 43\% of the galaxies in our representative sample are found to have a companion. 
All the galaxies in our representative sample with a companion are either classified as PRs or CKs, or show spatially resolved kinematic components. This correlation is suggestive of the importance of mergers and interactions on the kinematic properties of LCBGs.

Spatially resolved kinematic components are found in 14\% of the galaxies in our representative sample. 
While we find spatially resolved kinematic components that might be due to a minor merger in 10\% of the galaxies in our representative sample, evidence of an ongoing major merger is found only in one of our galaxies (5\% of the sample). 

Out of the 22 LCBGs in our representative sample, 21 (95\% of the sample) show no evidence of AGN activity. This suggests that such a phenomenon is not common in LCBGs.  UCM 0000 is the only galaxy with evidence for AGN activity.  In our data it shows both a spectrally resolved broad kinematic component ($\sigma\sim200$~km~s$^{-1}$) and emission line flux ratios characteristic of an AGN.

27\% of the galaxies in our representative sample show spectrally resolved kinematic components. Even though we cannot unambiguously state the nature of these components, both their intensities and offsets are in agreement with SN-driven galactic winds previously discussed by \citet{marlowe95} in a sample of dwarf galaxies with star formation activity. Nevertheless, SN-driven galactic winds do not seem to be typical among our representative sample of local LCBGs.

Galaxies in our sample classified as RDs show rotational velocities that range between $\sim50$ and $\sim200$~km~s$^{-1}$, which translate into dynamical masses that range between $\sim1\times10^9$ and $\sim3\times10^{10}$~$M_{\odot}$.

Those objects in our representative sample of 22 LCBGs which show rotating natures can be compared with the spiral galaxies used to calibrate the Tully-Fisher relation \citep{tully00}. A dispersion five times higher than the one found for spiral galaxies implies that the velocity widths of those LCBGs that rotate, rather than accounting exclusively for this rotation, may also include other kinematic components.

When compared, dynamical masses derived from rotation curves and integrated velocity widths differ in our representative sample of LCBGs up to a factor of $\sim4$. Such a difference may have an important impact on the study of distant LCBGs when observed through multi-object long-slit spectrographs. These kind of surveys rely on integrated spectral measurements to derive different physical properties, such as dynamical masses.

\section*{Acknowledgments}

It is a pleasure to thank the many people who welcomed our project into the 3.5-m telescope in CAHA, where we always felt at home. We also thank the invaluable work of our anonymous referee, which greatly improved the quality of this manuscript. J. P\'erez-Gallego thanks Sun Mi Chung, Bruno Ferreira, Carlos Hoyos, Carlos Rom\'an, and Sebasti\'an S\'anchez for their selfless help and interesting conversations throughout the development of the work presented in this paper. J. P\'erez-Gallego acknowledges support from a University of Florida Alumni Fellowship, and R. Guzm\'an from NASA grant LTSA NA65-11635. This work is partially funded by the Spanish MICINN under the Consolider Ingenio 2010 Program grant CSD2006-00070: First Science with the GTC (http://www.iac.es/consolider-ingenio-gtc/). This work is also partially funded by the Spanish Programa de Astronom\'ia y Astrof\'isica under grants AYA2006-02358 and AYA2006-15698-C02-02.

\appendix

\label{lastpage}

\end{document}